\documentclass[aps,twocolumn,pra,superscriptaddress,amsmath,showpacs,tightenlines]{revtex4}

\usepackage{amsmath}

\usepackage{graphicx,times}
\usepackage{amstext}
\usepackage{amsmath}            
\usepackage{amssymb}            
\usepackage{graphicx}           
\usepackage{latexsym}
\usepackage{bm}
 \usepackage{color}

 \def\extra#1{{#1}}

\def\Title#1{{#1}}

\def \etal{\textit{et al.}}

 \def\fig#1{{#1}}

\newcommand{\job}{J. Opt. B: Quantum Semiclass. Opt.~}

\def\tr{{\rm Tr}}

\def\EP{{\rm EP}}

\def\<{\langle}
\def\>{\rangle}

\newcommand{\ket}[1]{\mbox{$|#1\rangle$}}
\newcommand{\bra}[1]{\mbox{$\langle#1|$}}

\begin{document}

\title{Tunable multiphonon blockade in coupled nanomechanical resonators}




\author{Adam Miranowicz}
\affiliation{Faculty of Physics, Adam Mickiewicz University,
61-614 Pozna\'n, Poland} \affiliation{CEMS, RIKEN, Wako-shi,
Saitama 351-0198, Japan}

\author{Ji\v r\'\i\ Bajer}
\affiliation{Department of Optics, Palack\'{y} University, 772~00
Olomouc, Czech Republic}

\author{Neill Lambert}
\affiliation{CEMS, RIKEN, Wako-shi, Saitama 351-0198, Japan}

\author{Yu-xi Liu}
\affiliation{Institute of Microelectronics, Tsinghua University,
Beijing 100084, China} \affiliation{Tsinghua National Laboratory
for Information Science and Technology (TNList),  Beijing 100084,
China} \affiliation{CEMS, RIKEN, Wako-shi, Saitama 351-0198,
Japan}

\author{Franco Nori}
\affiliation{CEMS, RIKEN, Wako-shi, Saitama 351-0198, Japan}
\affiliation{Physics Department, The University of Michigan, Ann
Arbor, Michigan 48109-1040, USA}

\date{\today}

\begin{abstract}
A single phonon in a nonlinear nanomechanical resonator (NAMR) can
block the excitation of a second phonon [Phys. Rev. A {\bf 82},
032101 (2010)]. This intrinsically quantum effect is called phonon
blockade, and is an analog of Coulomb blockade and photon
blockade. Here we predict tunable multiphonon blockade in coupled
nonlinear NAMRs, where nonlinearity is induced by two-level
systems (TLSs) assuming dispersive (far off-resonance)
interactions. Specifically, we derive an effective Kerr-type
interaction in a hybrid system consisting of two
nonlinearly-interacting NAMRs coupled to two TLSs and driven by
classical fields. The interaction between a given NAMR and a TLS
is described by a Jaynes-Cummings-like model. We show that by
properly tuning the frequency of the driving fields one can induce
various types of phonon blockade, corresponding to the entangled
phonon states of either two qubits, qutrit and quartit, or two
qudits. Thus, a $k$-phonon Fock state (with $k=1,2,3$) can impede
the excitation of more phonons in a given NAMR, which we interpret
as a $k$-phonon blockade (or, equivalently, phonon tunneling). Our
results can be explained in terms of resonant transitions in the
Fock space and via phase-space interference using the
$s$-parametrized Cahill-Glauber quasiprobability distributions
including the Wigner function. We study the nonclassicality,
entanglement, and dimensionality of the blockaded phonon states
during both dynamics and in the stationary limits.
\end{abstract}

\pacs{42.50.Pq, 85.85.+j, 42.50.Dv}


\maketitle

\section{Introduction}

Nanomechanical and optomechanical devices are a versatile
technology~\cite{Huang03,Knobel03,Blencowe04,LaHaye04,
Blencowe04review,Schwab05review,Ekinci05review}, with a range of
applications in the quantum regime~\cite{Aspelmeyer14review}.
Examples include small mass or weak-force
detection~\cite{Caves80,Bocko96,Buks06}, quantum
measurements~\cite{Braginsky92}, and quantum-information
processing. The first success in putting a mechanical device in a
quantum state was performed~\cite{OConnell10} by purely cryogenic
means, due to the frequency of the mechanical
phonons~\cite{ClelandBook} being larger than the thermal energy.
Since then lower-frequency devices (which thus have larger mass)
have been put in quantum states using side-band cooling via
microwave and optical
cavities~\cite{Teufel11,Chan11,Safavi12,Verhagen12}. These
breakthroughs have been followed by the observation of state
transfer~\cite{Stannigel10,Palomaki13,Huang13,Fu14}  between an
electromagnetic cavity and the mechanical system, with the goal of
developing hybrid mechanical circuit
devices~\cite{Wallquist09,Stannigel10,Safavi11} for applications
in quantum technologies~\cite{Regal08,Woolley08}.

Moving into the nonlinear regime with such quantum nanomechanical
devices is desirable for several reasons: It will allow us to
observe highly nonclassical effects, as well as allowing to
individually address different transitions within the mechanical
system, so that it behaves as a mechanical artificial atom.  In
the realm of quantum
optics~\cite{Tian92,Leonski94,Miran96,Imamoglu97,Werner99,
Brecha99, Rebic99, Kim99, Rebic02, Smolyaninov02} and circuit
quantum electrodynamics~\cite{Hoffman11,Lang11,Liu14},
nonlinearities are typically associated with ``photon blockade''
(also referred to as the optical state truncation by quantum
nonlinear scissors~\cite{Miran01,Leonski11review}). In this regime
the nonlinear nature of the spectrum of an optical
cavity~\cite{Birnbaum05}, induced via, e.g., a Kerr nonlinearity,
means that the presence of a single photon within the  cavity
prevents the transmission of a second photon. In nonlinear
mechanical systems one expects an analogous ``phonon
blockade''~\cite{Liu10,Didier11} to arise. This requires either
strong intrinsic nonlinearities~\cite{Johansson14}, or induced
nonlinearity via ancillary nonlinear systems (like qubits or
artificial two-level systems~\cite{Buluta11}). Finally, while
coupling between mechanical phonons and electromagnetic photons
has been achieved in the quantum regime~\cite{Palomaki13b}, the
controlled coupling between multiple mechanical modes has so far
been restricted to classical devices~\cite{Okamoto13}. Such
controllable coupling would enable the observation of entangled
states~\cite{Cleland04,Armour02,Tian05}, and the Bell inequality
violation with massive objects~\cite{Johansson14}, as well as the
realization of coupled mechanical
qubits~\cite{Jacobs09,Hartman13}.

Our goal in this work is to study the combination of
nonlinearity-induced phonon blockade effects via the coupling
between the mechanical modes of NAMRs. We will show that the
infinite-dimensional mechanical states of the NAMRs, under proper
resonance conditions, can effectively be truncated to the states
of finite-dimensional systems of either two coupled qubits, a
three-level system (called a qutrit) coupled to a four-level
system (referred to as quartit or ququart), or, in general, two
coupled $d$-level systems (qudits).

While our model can be considered as a quantum limit of the
classical systems studied in~Ref.~\cite{Mahboob14}, we discuss
explicitly how the nonlinearities can be tuned using an ancilla
two-level system (TLS). We will show how this combination of
phonon-blockade and two-NAMR (or two-mode) coupling leads to
multiphonon blockade (or phonon tunneling), in analogy to the
predictions of multiphoton
blockade~\cite{Miran13,Hovsepyan14,Miran14qre,Wang15} and closely
related photon
tunneling~\cite{Smolyaninov02,Faraon08,Majumdar12,Wang15}. For
example, Ref.~\cite{Smolyaninov02} provides a pedagogical
explanation of how photon blockade can lead to the observation of
a single-photon tunneling effect in an analogous way to how
Coulomb blockade can lead to the observation of single-electron
tunneling.

We discuss here how the form of the multiphonon blockade can be
tuned via driving, and verify the resultant highly nonclassical
states with a variety of measures.

This paper is organized as follows: In Sec.~II, we describe a
model for a hybrid system of coupled linear NAMRs and TLSs. In
this section and in Appendix~A, we also show how a Kerr-type
nonlinearity can be induced via NAMR-TLS interactions, and derive
an effective Hamiltonian for the coupled nonlinear NAMRs. The
possibility of observing multiphonon blockades in this system is
described in Sec.~III. We summarize several methods to assess
nonclassicality, which we then apply in our analysis of phonon
blockade in Secs.~IV and V. We conclude in Sec.~VI.

\section{Model}

\begin{figure} 
\includegraphics[width=8cm]{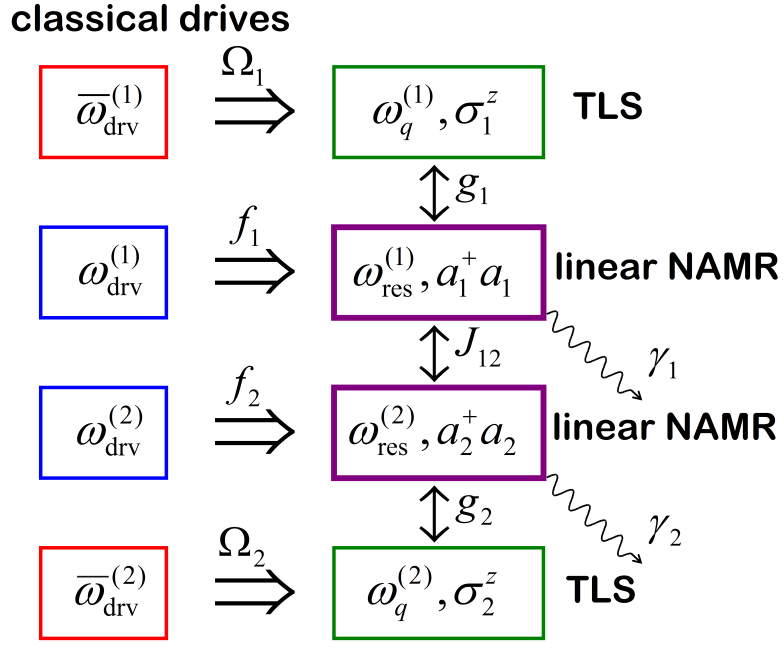}
\caption{(Color online) Schematic diagram for the couplings in the
dissipative hybrid system described by the Hamiltonian given in
Eq.~(\ref{N0}). The system consists of two linear nanomechanical
resonators (NAMRs), with frequencies $\omega_{\rm res}^{(n)}$, and
a pair of two-level systems (TLSs), with frequencies $\omega_{\rm
q}^{(n)}$, driven by classical fields, with frequencies
$\omega_{\rm drv}^{(n)}$ and $\bar \omega_{\rm drv}^{(n)}$,
respectively. Moreover, $\Omega_n$, $f_n$, $g_n$, and $J_{12}$
denote the coupling strengths of the depicted subsystems;
$\gamma_n$ are the NAMR decay rates; $a_n$ ($a_n^\dagger$) is the
phonon annihilation (creation) operator for the $n$th NAMR, and
$\sigma_n^{z}$ is the Pauli operator for the $n$th TLS. }
\label{fig01}
\end{figure}

\subsection{Time-dependent Hamiltonian}

We consider a hybrid system, as schematically depicted in
Fig.~\ref{fig01}, consisting of two interacting driven linear
nanomechanical resonators (NAMRs), described by the Hamiltonian
$H_{\rm res}$, coupled to two driven two-level systems (TLSs,
qubits), given by the Hamiltonian $H_{\rm q}$. The interaction
$H_{\rm JC}$ between the $n$th NAMR and $n$th TLS (for $n=1,2$) is
described by a Jaynes-Cummings-like model under the rotating
wave-approximation. The interaction $H_{\rm int}$ between the two
NAMRs can be interpreted as a combined driven process of frequency
conversion and parametric amplification~\cite{Mahboob14}. Thus,
the total microscopic Hamiltonian, representing the system shown
in Fig.~\ref{fig01}, reads (hereafter $\hbar=1$ and $n=1,2$):
\begin{eqnarray}
H & =& H_{\rm res} + H_{\rm q} + H_{\rm JC} + H_{\rm int},
\label{N0} \\
H_{\rm res}&=&\sum_{n}\omega_{\rm res}^{(n)}a_{n}^{\dagger}a_{n}
+f_{n}\big[a_n\exp(i\omega^{(n)}_{\rm drv}t)+{\rm H.c.}\big]
\label{N1} ,\\
H_{\rm q}&=&\sum_{n}\frac{\omega_{\rm q}^{(n)}}{2}\sigma_n^{z}
+\frac{\Omega_{n}}{2}\big[\sigma^{-}_{n}\exp(i\overline\omega^{(n)}_{\rm
drv}t)+{\rm H.c.}\big]
\label{N2} ,\\
H_{\rm JC}&=&\sum_{n}g_{n}(a_{n}\sigma^{+}_{n}
+a_{n}^{\dagger}\sigma^{-}_{n} ),
\label{N3} \\
H_{\rm int}&=&J_{12}\big[a_1\exp(i\omega^{(1)}_{\rm drv}t)+{\rm
H.c.}\big]\nonumber\\ &&\qquad
\times\big[a_2\exp(i\omega^{(2)}_{\rm drv}t)+{\rm H.c.}\big],
\label{N4}
\end{eqnarray}
where $a_n$ and $a_n^\dagger$ are, respectively, the phonon
annihilation and creation operators for the $n$th NAMR,
\begin{eqnarray}
a_n=(2m_n \omega_{\rm res}^{(n)})^{-1/2} (m_n\omega_{\rm
res}^{(n)}x_n +ip_n),
\nonumber \\
a_n^\dagger=(2m_n \omega_{\rm res}^{(n)})^{-1/2} (m_n\omega_{\rm
res}^{(n)}x_n -ip_n),
 \label{N48}
\end{eqnarray}
which are given in terms of the position operator $x_n$, momentum
operator $p_n$, and frequency $\omega_{\rm res}^{(n)}$ of  the
NAMR. Moreover,
$\sigma_n^{z}=\ket{e_n}\bra{e_n}-\ket{g_n}\bra{g_n}$ is the Pauli
$Z$ operator for the $n$th TLS, while
$\sigma_n^-=\ket{g_n}\bra{e_n}$ ($\sigma_n^+=\ket{e_n}\bra{g_n}$)
is the qubit lowering (raising) operator given in terms of the
ground ($\ket{g_n}$) and excited ($\ket{e_n}$) states of the $n$th
TLS; $\omega_{\rm q}^{(n)}$ is the TLS frequency and $\omega_{\rm
drv}^{(n)}$ ($\overline \omega_{\rm drv}^{(n)}$) is the NAMR (TLS)
driving-field frequency. The coupling strengths of the subsystems,
as shown in Fig.~\ref{fig01}, are denoted by $\Omega_n$, $f_n$,
$g_n$, and $J_{12}$. The symbol H.c. denotes the Hermitian
conjugated term.

This system, described by Eq.~(\ref{N0}), can be realized in
various ways as a combination of two types of implementations,
e.g.: (i) the proposal of Ref.~\cite{Liu10} for observing
single-mode phonon blockade in a driven single NAMR coupled to a
superconducting quantum two-level system and (ii) the system of
two nonlinearly-coupled NAMRs, which was experimentally realized
in the NTT experiments (see, e.g., Ref~\cite{Mahboob14} and
references therein). It is worth noting that we assumed that the
interaction $H_{\rm int}$ is additionally driven at frequencies
$\omega^{(1)}_{\rm drv}$  and $\omega^{(2)}_{\rm drv}$, as given
in Eq.~(\ref{N4}), which slightly generalizes the model applied in
Ref.~\cite{Mahboob14}. We also note that the interaction described
by Eq.~(\ref{N3}) conserves the number of excitations, in contrast
to that described by Eq.~(\ref{N4}).

\subsection{Time-independent Hamiltonian in a rotated dressed-qubit
basis}

In the following, we assume  the following large detunings:
\begin{eqnarray}
\Delta_{\rm rq}^{(n)}&\equiv&\omega_{\rm
res}^{(n)}-\omega^{(n)}_{\rm q}=\omega_{\rm
drv}^{(n)}-\overline\omega^{(n)}_{\rm drv}\gg g_n >0,
 \label{detuning1} \\
\Delta_{n}&\equiv&\Delta_{\rm rd}^{(n)}\equiv\omega_{\rm
res}^{(n)}-\omega^{(n)}_{\rm drv}=\omega_{\rm
q}^{(n)}-\overline\omega^{(n)}_{\rm drv}\gg \Omega_n >0.\qquad
 \label{detuning2}
\end{eqnarray}
The large detuning $\Delta_{\rm rq}^{(n)}\gg g_n$ implies that,
e.g., the qubit states cannot be flipped by the interaction with
the NAMR (see Appendix~A). While the large detuning $\Delta_{n}\gg
\Omega_n$ enables us to omit, in particular, the terms which do
not conserve the number of excitations in the TLS-NAMR interaction
Hamiltonian, as will be explained below. Note that the assumption
that $\omega_{\rm res}^{(n)}-\omega^{(n)}_{\rm drv}=\omega_{\rm
q}^{(n)}-\overline\omega^{(n)}_{\rm drv}$ is not essential in our
derivation, and is applied only for simplicity.

First, we transform the Hamiltonian $H$, given in Eq.~(\ref{N0}),
into a rotating reference frame by the unitary transformation
\begin{equation}
  U_R(t)=\prod_n\exp\left(-i\omega^{(n)}_{\rm
drv}a_{n}^{\dagger}a_{n}t
  -\tfrac12 i\bar\omega^{(n)}_{\rm
drv}\sigma_n^{z}t   \right),
 \label{N6}
\end{equation}
which results in the following effective Hamiltonian
\begin{eqnarray}
  H'&=&U_R^\dagger H U_R -i U_R^\dagger \frac{\partial}{\partial t}
  U_R, \label{Hprime}
\end{eqnarray}
which can be given explicitly as
\begin{eqnarray}
H' &=& H'_{\rm res} + H'_{\rm q} + H'_{\rm JC} + H'_{\rm int},
\label{Hprim}\\
H'_{\rm res}&=&\sum_{n}\Delta_{n}a_{n}^{\dagger}a_{n}
+f_{n}\big(a_n+a_{n}^{\dagger}\big)
\label{N9} ,\\
H'_{\rm q}&=&\sum_{n}\frac{\Delta_{n}}{2}\sigma_n^{z}
+\frac{\Omega_{n}}{2}\sigma_n^{x}
\label{N10a} ,\\
 H'_{\rm JC}&=&\sum_{n}g_{n}[a_{n}\sigma^{+}_{n}
\exp(-i\Delta^{(n)}_{\rm rq}t)+{\rm H.c.}],
\label{H_JC1} \\
H'_{\rm int}&=& J_{12}(a_{1}
+a_{1}^{\dagger})(a_{2}+a_{2}^{\dagger} ), \label{Hint1}
\end{eqnarray}
where $\sigma_n^{x}=\sigma^{+}_{n}+\sigma^{-}_{n}$. Note that this
Hamiltonian is still time dependent.

Now we diagonalize the qubit Hamiltonian, given by
Eq.~(\ref{N10a}), by transforming it into a dressed-qubit basis
following the method described in, e.g.,
Refs.~\cite{CohenBook,Liu06}. Thus, one finds
\begin{eqnarray}
  H''_{\rm q}=\sum_{n}\frac{\bar\Delta_{n}}{2}
\rho^{z}_{n} \label{N10} ,
\end{eqnarray}
where $\bar\Delta_{n}=\sqrt{\Delta_{n}^2+\Omega^2_{n}}$ and the
dressed-qubit operator
$\rho_n^z=\ket{E_n}\bra{E_n}-\ket{G_n}\bra{G_n}$ can be defined by
the dressed-qubit basis states~\cite{Liu06}:
\begin{eqnarray}
  \ket{E_n} &=& \cos x_n\ket{e_n}+\sin x_n\ket{g_n},
\nonumber \\
  \ket{G_n}&=&-\sin x_n\ket{e_n}+\cos x_n\ket{g_n},
\label{EG}
\end{eqnarray}
where $x_n=(1/2)\tan^{-1}(\Omega_{n}/\Delta_{n})$. It is seen that
the dressed $n$th qubit refers to the $n$th TLS dressed with the
$n$th NAMR phononic field (for a general discussion see
Ref.~\cite{CohenBook}).

To transform the Hamiltonian $H'_{\rm JC}$, given by
Eq.~(\ref{H_JC1}), into the dressed-qubit basis, first we note
that
\begin{equation}
  a\sigma_n^+
   = \cos^2(x_n)a\rho^+_n-\sin^2(x_n)a\rho^-_n+\tfrac12
  \sin(2x_n)a\rho_n^z,
\label{dressed1}
\end{equation}
where $\rho_n^-=\ket{G_n}\bra{E_n}$, and
$\rho_n^+=\ket{E_n}\bra{G_n}$. By recalling the assumption, given
in Eq.~(\ref{detuning2}), we can write $a\sigma_n^+ \approx
\cos^2(x_n)a\rho^+_n$. For example, if $\Omega_{n}/\Delta_{n}=0.1$
then $\cos^2(x_n)=0.9975$, $\sin^2(x_n)=0.0025$, and
$\sin(2x_n)/2\approx 0.05$. Thus, we can omit the second and third
terms in Eq.~(\ref{dressed1}), and their Hermitian conjugates,
which do not conserve the number of excitations. Then, the
Hamiltonian $H'_{\rm JC}$ can approximately be transformed into
\begin{equation}
H''_{\rm JC}\approx \sum_{n}g'_{n}[a_{n}\rho^{+}_{n}
\exp(-i\Delta^{(n)}_{\rm rq}t)+{\rm H.c.}], \label{H_JC2}
\end{equation}
where $g'_n=g_n\cos^2(x_n)$.

To obtain a time-independent total Hamiltonian, we transform it
into a qubit rotating frame by applying the standard unitary
transformation
\begin{equation}
  U_q=\exp(-i H''_q t).
 \label{U2}
\end{equation}
This results in
\begin{equation}
H'''_{\rm JC}= \sum_{n}g'_{n}\left\{a_{n}\rho^{+}_{n}
\exp[-i(\Delta^{(n)}_{\rm rq}-\bar\Delta_n)t]+{\rm H.c.}\right\}.
\label{H_JC3detuning}
\end{equation}
By assuming $\Delta^{(n)}_{\rm rq}=\bar\Delta_n$ (for $n=1,2$),
one obtains the time-independent Jaynes-Cummings Hamiltonian in
the dressed-qubit basis,
\begin{equation}
H'''_{\rm JC}= \sum_{n}g'_{n}(a_{n}\rho^{+}_{n} +
a^\dagger_{n}\rho^{-}). \label{H_JC3}
\end{equation}
Thus, after these transformations, the total Hamiltonian reads
\begin{equation}
  H''' = H'''_{\rm res}  + H'''_{\rm JC} + H'''_{\rm int},
\label{H3}
\end{equation}
where $H'''_{\rm int}=H'_{\rm int}$ and  $H'''_{\rm res}=H'_{\rm
res}$. Note that $H'''_q=0$.

\begin{figure} 
\includegraphics[width=8cm]{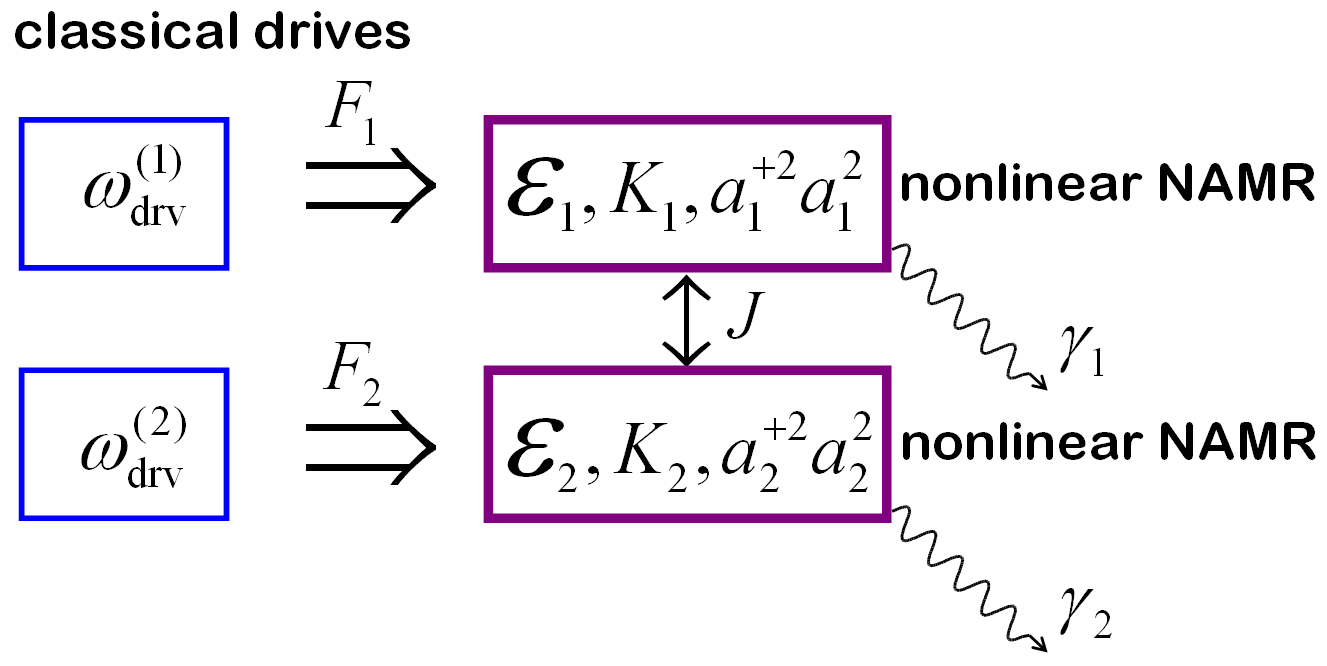}
\caption{(Color online) Schematic diagram for the couplings in the
dissipative hybrid system described by the effective Hamiltonian
given in Eq.~(\ref{Heff}), which consists of two nonlinear NAMRs
driven by classical fields. Here $F_n$ and $J$ denote the
corresponding coupling strengths; $K_n$ is the effective Kerr
nonlinearity of the $n$th nonlinear NAMR, and ${\cal E}_n$ is its
energy. Other symbols are the same as in Fig.~\ref{fig01}.}
\label{fig02}
\end{figure}
\subsection{Effective Hamiltonian with a qubit-induced
nonlinearity}

We assume another large detuning
\begin{eqnarray}
\delta_n&\equiv&\bar\Delta_{n}-\Delta_{n}\gg g'_n,
\label{detuning3}
\end{eqnarray}
which leads to dispersive interactions. Note that
conditions~(\ref{detuning2}) and~(\ref{detuning3}) are not
contradictory as can be shown as follows: By denoting
$r_n=\Omega_n/\Delta_{n}$, Eq.~(\ref{detuning3}) can be given as
\begin{eqnarray}
\delta_n=\Delta_{n}\left(\sqrt{1+r_n^2}-1\right)\approx \tfrac12
\Delta_{n}r_n^2 \gg g'_n. \label{detuning3a}
\end{eqnarray}
Thus,  Eqs.~(\ref{detuning2}) and~(\ref{detuning3}) can be
combined as the following hierarchy of conditions:
\begin{eqnarray}
\Delta_{n}\gg \Omega_n \gg g'_n,\quad {\rm such~that}\quad \Omega_n^2
\gg 2g'_n\Delta_n. \label{detuning23}
\end{eqnarray}
For example, if $r_n=0.1$ then we require $0.005\Delta_n\gg g_n'$.

\begin{figure} 
\includegraphics[width=8cm]{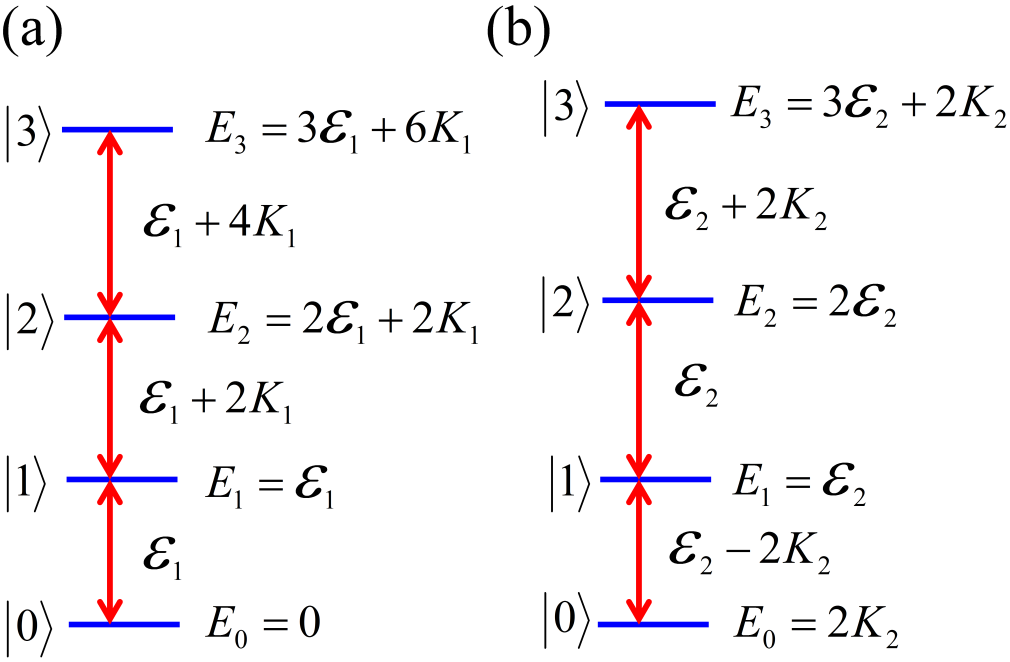}
\caption{(Color online) Energy levels for the Kerr-type
Hamiltonians given in (a) Eq.~(\ref{Hkerr1}) (with $n=1$) and (b)
Eq.~(\ref{Hkerr2}) (with $n=2$ and ${\cal E}_2\equiv {\cal E}_2
^{12}$).} \label{fig03}
\end{figure}

Now we describe how the TLS-NAMR interaction can effectively
induce a Kerr-type nonlinearity. To show this, we can expand the
Hamiltonian $H'$ in a power series of the parameter
\begin{equation}
  \lambda_{n} =\frac{g'_n}{\delta_n}=
  \frac{g_n\cos^2\big[\tfrac12\tan^{-1}(\Omega_{n}/\Delta^{}_{n})]}
  {\sqrt{\Delta_{n}^2+\Omega^2_{n}}-\Delta_{n}},
 \label{lambda}
\end{equation}
such that $|\lambda_n|\ll 1$. As derived in Appendix~A, one can
keep terms of such expansions up to $\lambda_n^3$ only and assume
that both TLSs remain in their excited states $\ket{E_n}$ during
the whole system evolution, which can be observed for the large
detuning $\Delta_{\rm rq}^{(n)}\gg g_n$, as given in
Eq.~(\ref{detuning1}). Then, the total Hamiltonian $H'''$, given
in Eq.~(\ref{H3}), can be transformed into the following effective
Hamiltonian:
\begin{eqnarray}
H_{\rm eff} & =& \sum_{n}\Big[H^{(n)}_{\rm Kerr}(0,1)-\omega_{\rm
drv}^{(n)} a_n^{\dagger}a_n
 +F_n\left(a_n+a_n^{\dagger}\right)\Big]
\nonumber \\
&&\quad+J(a_{1}  +a_{1}^{\dagger})(a_{2}+a_{2}^{\dagger} ),
\label{Heff}
\end{eqnarray}
where the Kerr-type Hamiltonian
\begin{eqnarray}
H^{(n)}_{{\rm Kerr}}(0,1) & = & K_n a_n^{\dagger}a_n^{\dagger}a_n
a_n+{\cal E}_n ^{} a_n^{\dagger}a_n \label{Hkerr1}
\end{eqnarray}
describes an effectively nonlinear $n$th NAMR. Here the $n$th NAMR
energy ${\cal E}_n$ and the effective Kerr nonlinearity $K_n$ are
given by
\begin{eqnarray}
{\cal E}_n &=& \omega_{\rm res}^{(n)}+2K_n+g'_n\lambda_n
(1-\lambda_n^2),
\\
K_n&=&-g'_n\lambda^3_n,
 \label{Kerr}
\end{eqnarray}
respectively. Moreover, $F_n=f_{n}(1+\tfrac12 \lambda_n^2)$ is an
effective driving-field strength, and
$J=J_{12}(1+\tfrac12\lambda_1^2)(1+\tfrac12\lambda_2^2)$ is an
effective coupling between the NAMRs. All these coupling
coefficients are shown in Fig.~\ref{fig02}. A few lowest energy
levels for the Kerr-type part of this Hamiltonian, given in
Eq.~(\ref{Hkerr1}), are shown in Fig.~\ref{fig03}(a), where we set
$n=1$.

Equation~(\ref{Heff}) describes an effective driven Kerr-type
self-interaction (an anharmonic model) in one phonon mode
nonlinearly coupled to another phonon mode.  In the context of
various types of photon and phonon blockades, this phonon-phonon
model can formally be considered as a two-mode generalization of
the single-mode phonon model of Ref.~\cite{Liu10}. Moreover, our
model with the nonlinear coupling between the NAMRs can be
interpreted as a generalization of (i) the linearly-coupled hybrid
model studied in Ref.~\cite{Didier11} to describe single-phonon
and single-photon blockades and (ii) the linearly-coupled optical
model with single-mode~\cite{Leonski04coupler,Liew10,Bamba11} and
two-mode drivings~\cite{Miran06coupler,Xu14coupler} leading to
two-mode single-photon blockades.

The Kerr Hamiltonian, given in Eq.~(\ref{Hkerr1}) can formally be
rewritten as ($k,l=0,1,...$)
\begin{equation}
H^{(n)}_{{\rm Kerr}}(k,l)  =  K_n
(a_n^{\dagger}a_n-k)(a_n^{\dagger} a_n-l)+{\cal E}_n ^{kl}
a_n^{\dagger}a_n -C_{n}^{kl}, \label{Hkerr_kl}
\end{equation}
where ${\cal E}_n ^{kl}={\cal E}_n +(k+l-1)K_n$, and
$C_n^{kl}=klK_n$ is a constant term, which can be ignored. Thus,
Eq.~(\ref{Hkerr_kl}) corresponds to $H^{(n)}_{{\rm Kerr}}(0,1)$. In
the following we will also analyze another special case of
Eq.~(\ref{Hkerr_kl}) corresponding to
\begin{equation}
  H^{(n)}_{{\rm Kerr}}(1,2)\cong K_n (a_n^{\dagger} a_n -1)
  (a_n^{\dagger} a_n -2) + {\cal E}_n ^{12}a_n^{\dagger} a_n,\label{Hkerr2}
\end{equation}
where, for simplicity, the term $C_{n}^{12}$ is omitted. A few
lowest energy levels for this Kerr Hamiltonian are shown in
Fig.~\ref{fig03}(b), where $n=2$ and ${\cal E}_2\equiv {\cal E}_2
^{12}$.

The driving of qubits, as given in the Hamiltonian~(\ref{N2}), can
tune the effective Kerr nonlinearity. In addition, as will be
discussed in the following and was also observed in
Refs.~\cite{Liew10,Bamba11}, there is another mechanism for tuning
the Kerr nonlinearity in the coupled anharmonic oscillators. Thus,
even if the NAMR decay rates $\gamma_{n}$ are much larger than the
NAMR driving strengths $F_{n}$ and the latter are much larger than
the Kerr nonlinearities $K_{n}$, strong single-time photon (or
phonon) antibunching can still be observed as an indicator of the
photon (phonon) blockade~\cite{Liew10,Bamba11}.

We will analyze free and dissipative evolutions of the NAMR
systems described by the Hamiltonian given in Eq.~(\ref{Heff}),
under two different resonance conditions, as specified later in
Eqs.~(\ref{Heff1}) and~(\ref{Heff2}).

\section{Phonon blockade}

\subsection{Phonon blockade in two models}

In this section we will analyze phonon blockade in two models,
which are special cases of the general model described by $H_{{\rm
eff}}$ in Eq.~(\ref{Heff}) under different resonance conditions.

\begin{figure} 
\includegraphics[height=5cm]{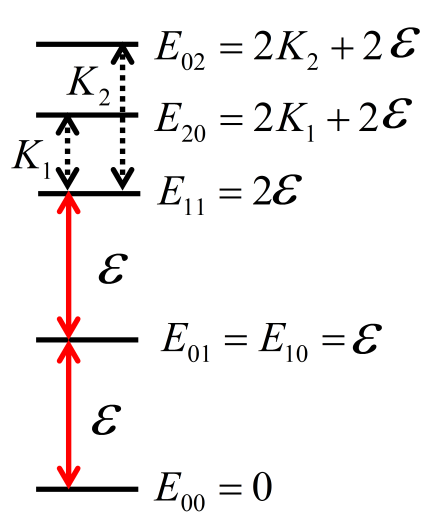}
\caption{(Color online) Energy levels $E_{m_1m_2}$, given by $\bar
H\ket{m_1,m_2}=E_{m_1m_2}\ket{m_1,m_2}$, for a simplified model~2,
i.e., the system of two uncoupled nonlinear NAMRs,  described by
the Hamiltonian $\bar H=\sum_n H^{(n)}_{\rm Kerr}$, given in
Eq.~(\ref{Hkerr1}) assuming ${\cal E}\equiv{\cal E}_1={\cal E}_2$.
Here, $m_{1}$ and $m_{2}$ are the Fock states of the two NAMRs. It
is seen that the levels $E_{20}$ and $E_{02}$ are off-resonance if
$K_1,K_2\neq {\cal E}/2$ and, thus, they are much less occupied
than the other shown levels. This explains the occurrence of
phonon blockade in this coupled system.} \label{fig04}
\end{figure}

In one model, we assume that the frequency $\omega_{\rm
drv}^{(n)}$ of the driving field of the $n$th NAMR is tuned
precisely to the effective free energy ${\cal E}_n$ ($n=1,2$).
Then, the effective Hamiltonian $H_{\rm eff}$, given in
Eq.~(\ref{Heff}),  simplifies to
\begin{eqnarray}
H'_{\rm eff} & = & \sum_nK_n a_n^{\dagger}a_n(a_n^{\dagger}a_n-1)
\notag \\
&&+\sum_nF_n\left(a_n+a_n^{\dagger}\right) +J(a_{1}
+a_{1}^{\dagger} )(a_{2} +a_{2}^{\dagger} ),\quad  \label{Heff1}
\quad\end{eqnarray} which is referred here to as \emph{model~1}.
The occurrence of single-phonon blockade in this model is
explained in Fig.~\ref{fig04}.

\begin{figure} 
\includegraphics[height=4cm]{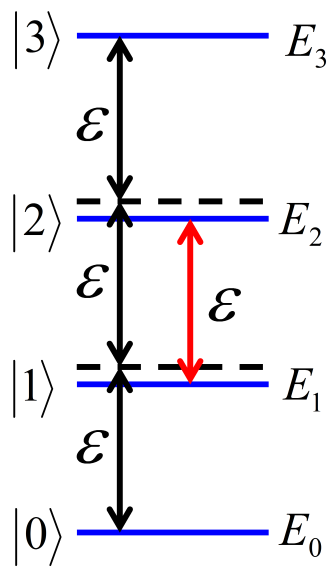}
\caption{(Color online) Energy levels for a single NAMR, as a  simplified
version of model~2: How to induce a three-phonon resonant
transition in a single nonlinear NAMR described by the Hamiltonian
given in Eq.~(\ref{Hkerr2}). If the driving field frequency is
resonant with the transition between the energy levels $\ket{1}$
and $\ket{2}$, $E_2-E_1={\cal E}$, then one can also induce the
three-phonon transition between the energy levels $\ket{0}$ and
$\ket{3}$, since $E_3-E_0=3{\cal E}$. Solid (dashed) lines denote
real (virtual) energy levels. We note that only single-phonon
transitions can be observed if the driving field frequency is
tuned with the transition between other levels $\ket{n}$ and
$\ket{n+1}$ ($n\neq 1$). Moreover, we cannot observe such
multiphonon transitions if the system is described by the
Hamiltonian given in Eq.~(\ref{Hkerr1}) for any $n$.}
\label{fig05}
\end{figure}

\begin{figure} 
\includegraphics[width=4cm]{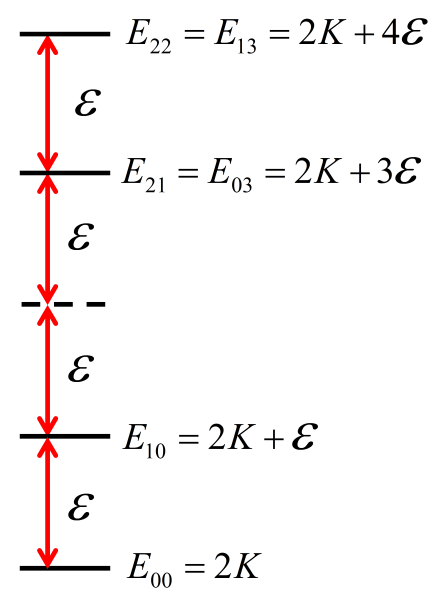}

\caption{(Color online) Energy levels for a simplified model~2,
i.e., for the NAMR system described by the Hamiltonian $H''_{\rm
eff}$, given in Eq.~(\ref{Heff2}), assuming two drives and the
same Kerr nonlinearities $K_1=K_2\equiv K$ and free energies
${\cal E}_1={\cal E}_2\equiv {\cal E}$ of both NAMRs, and setting
$J=F_1=F_2=0$. This graph explains the occurrence of multiphonon
blockade in this coupled system. The occurrence of the resonant
three-phonon transition is explained in Fig.~\ref{fig05}. We note
that, in particular, the following levels are off-resonance:
$E_{01}={\cal E}\neq E_{10}$, $E_{12}=3{\cal E}\neq E_{21}$,
$E_{02}=E_{11}=2{\cal E}\neq E_{10}+{\cal E}$, and $E_{20}=2{\cal
E}+4K\neq E_{10}+{\cal E}$.} \label{fig06}
\end{figure}

In another model, we set the frequency $\omega_{\rm drv}^{(n)}$ of
the driving field of the first (second) NAMR to be tuned with the
effective free energy ${\cal E}_1''={\cal E}_1$ (${\cal
E}_2''={\cal E}_2+2K_2$). Thus, the effective Hamiltonian $H_{\rm
eff}$ reduces to:
\begin{eqnarray}
H_{\rm eff}'' & = & K_1 a_1^{\dagger}a_1 (a_1^{\dagger} a_1-1)
 + K_2 (a_2^{\dagger} a_2-1) (a_2^{\dagger} a_2-2)
\nonumber \\
&&+\sum_n F_n\left(a_n+a_n^{\dagger}\right) +J(a_{1}
+a_{1}^{\dagger} )(a_{2} +a_{2}^{\dagger} ),\qquad \label{Heff2}
\end{eqnarray}
which is referred here to as \emph{model~2}. Note that we ignored
in Eq.~(\ref{Heff2}) the irrelevant constant terms $C_n^{kl}$ (for
$n=1,2$). Some lowest energy levels for the Kerr-type Hamiltonian
of Eq.~(\ref{Heff2}) for $n=1$ are shown in Fig.~\ref{fig03}(a) in
comparison with those for $n=2$ shown in Fig.~\ref{fig03}(b). A
closer analysis of Fig.~\ref{fig03}(b) shows that $E_3-E_0=3{\cal
E}_2\equiv 3{\cal E}$, which implies that a three-phonon resonant
transition can be observed, as shown in Fig.~\ref{fig05}. The
occurrence of multiphonon blockade in this coupled system can
clearly be understood by analyzing Fig.~\ref{fig06}.

\subsection{Phonon blockade in non-dissipative systems}

\begin{figure} 
\includegraphics[width=4.25cm]{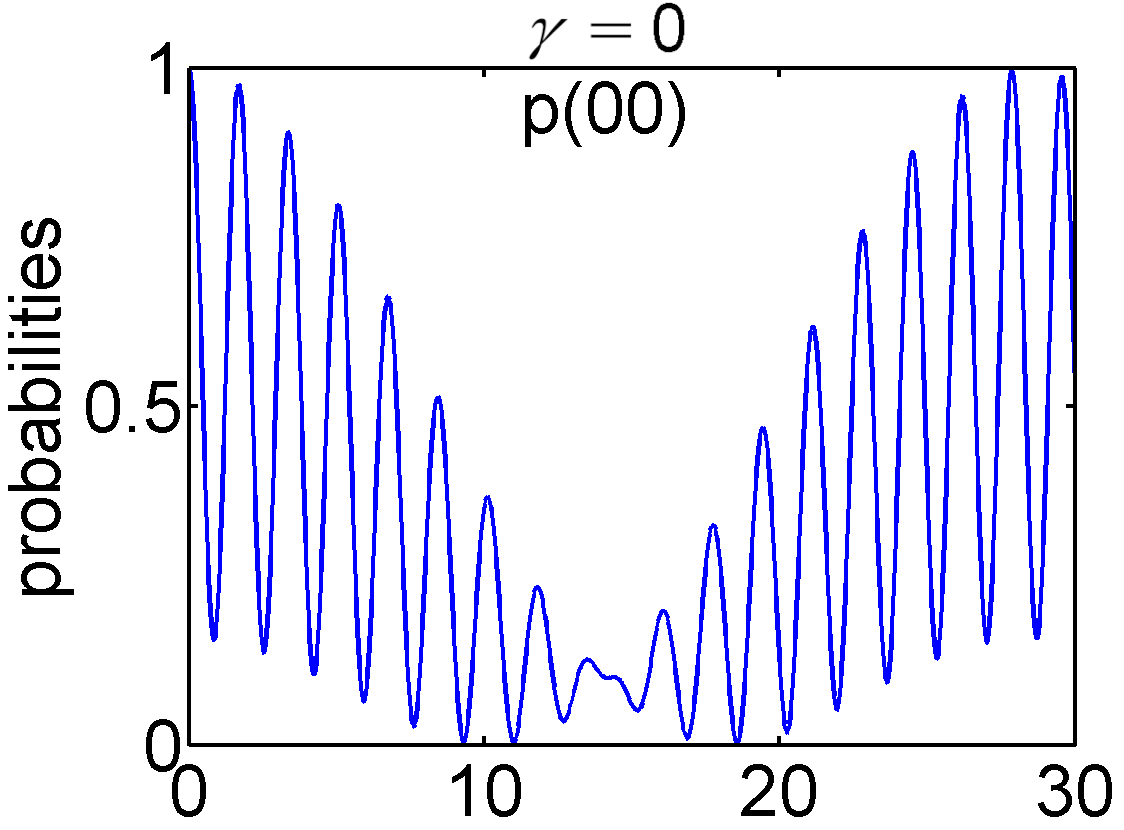}
\includegraphics[width=4cm]{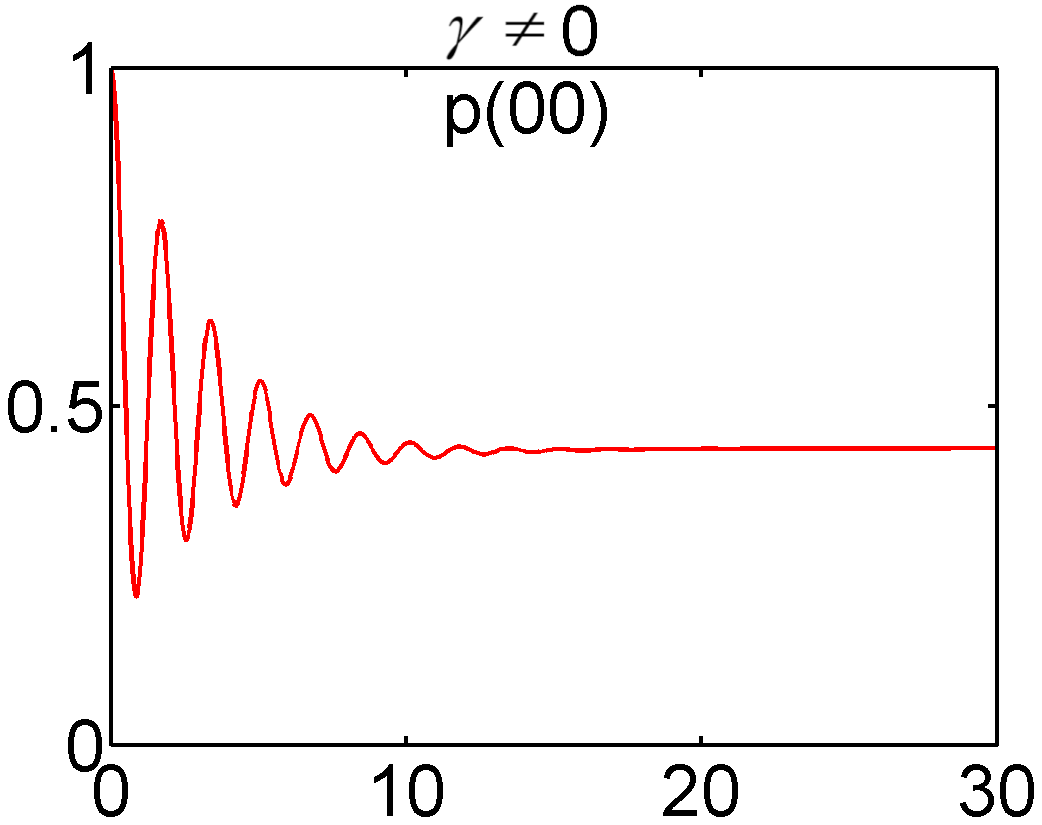}

\includegraphics[width=4.25cm]{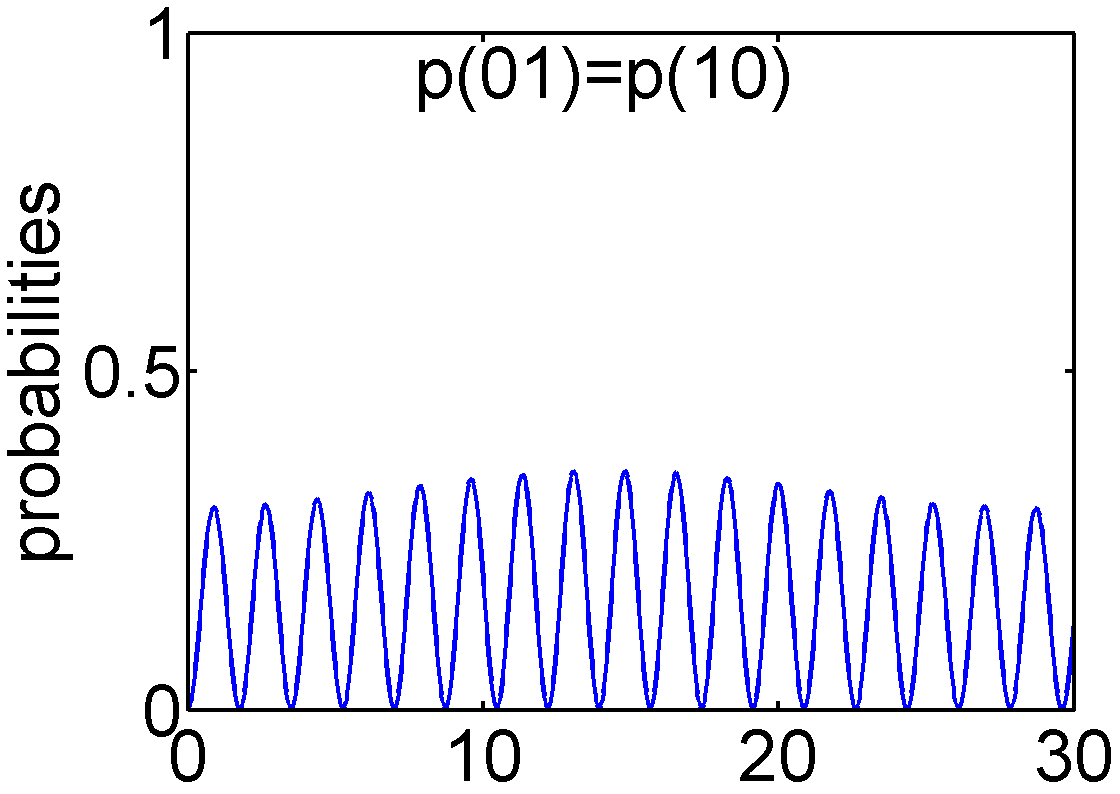}
\includegraphics[width=4cm]{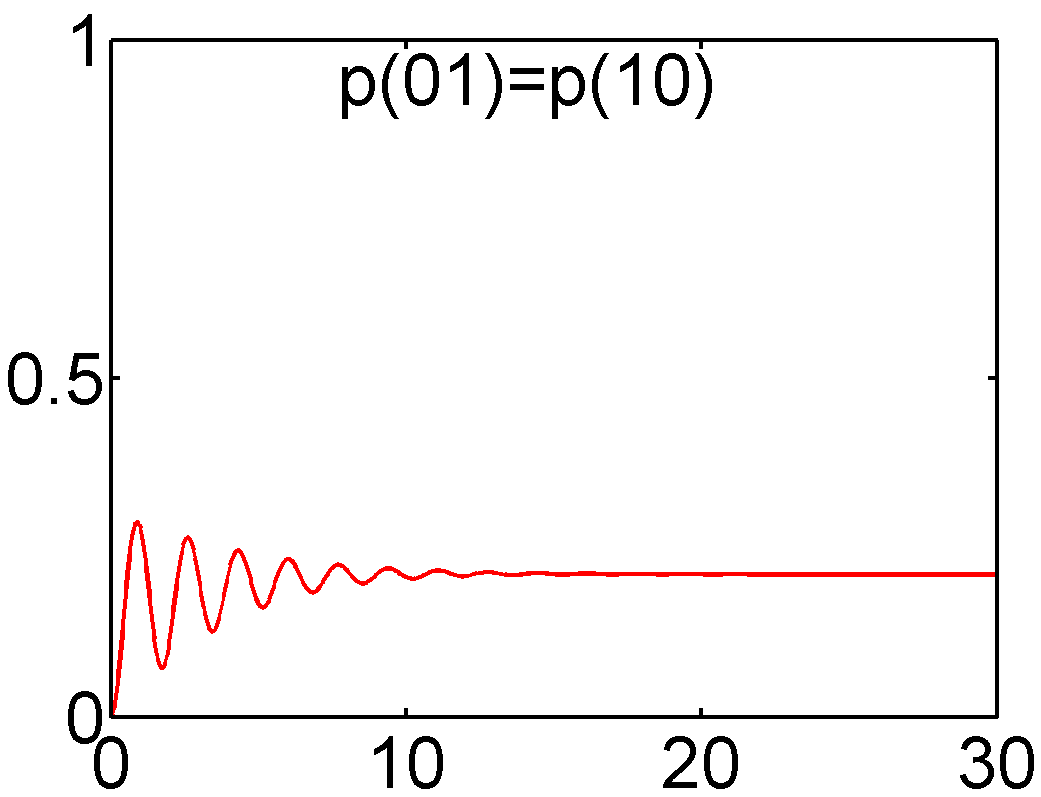}

\includegraphics[width=4.28cm]{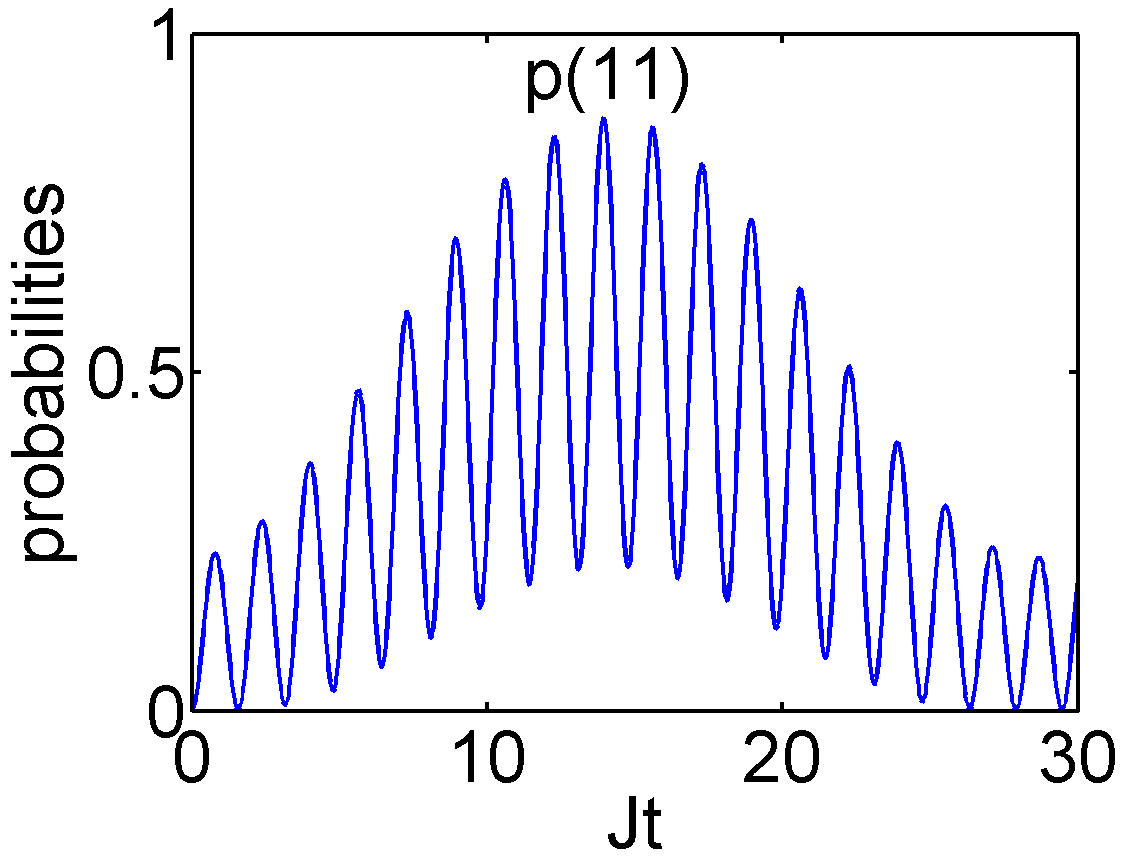}
\includegraphics[width=4cm]{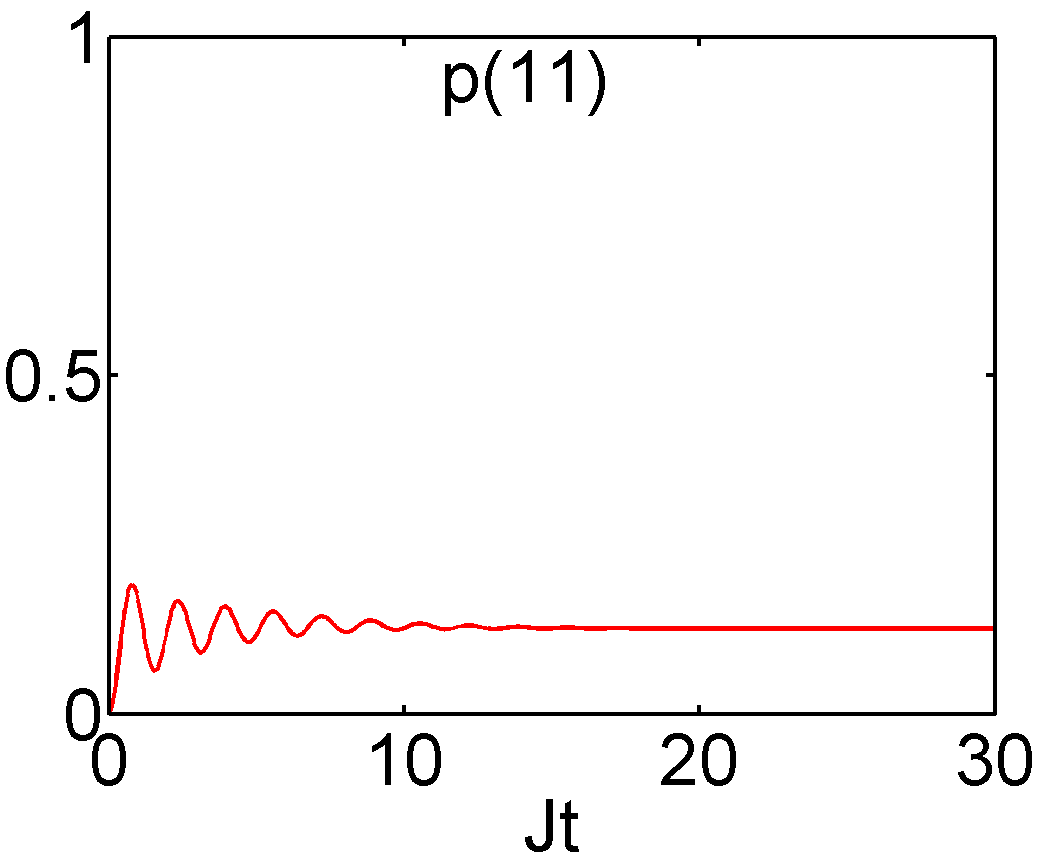}

\caption{(Color online) Non-stationary phonon blockade in model~1:
Evolutions of probabilities $P(m_1,m_2)=|c_{m_1,m_2}|^2$ for the
non-dissipative (left frames, $\gamma=0$) and dissipative (right
frames, $\gamma=J/3$) systems of the two nonlinear NAMRs described
by the effective Hamiltonian $H'_{\rm eff}$ plotted for the
rescaled time $Jt$, where $J$ is the interaction strength between
the NAMRs. Probabilities $P(m_1,m_2)$ for the other values of
$m_1,m_2$ are negligible on the scale of these figures and, thus,
not presented here. We set the Kerr nonlinearities $K_n=10J$, the
drive strengths $F_n=J$, and the mean number of thermal phonons
$\bar{n}_{\rm th}^{(n)}=0.01$ (right frames), for $n=1,2$. It is
seen in the right frames that the oscillations are rapidly damped.
Surprisingly, the stationary damped states are highly nonclassical
as it is shown in other figures.} \label{fig07}
\end{figure}

In order to show phonon blockade in the non-dissipative model~1,
we start from the analysis of the system, described by the
Hamiltonians $H'_{\rm eff}$, without dissipation. Hereafter we
assume that both NAMRs were initially in the ground phonon states,
$|\psi (t=0)\rangle =| 0 0\rangle$. A generalization of such
solution for other initial states is simple. For simplicity, we
assume that both NAMRs are driven equally with the strength
$F_1=F_2\equiv F$. Then, under these assumptions, the solution of
the Schr\"odinger equation for the wave function
$\ket{\psi}=\exp(-iH'_{\rm eff}t)\ket{00}$ is given by
\begin{equation}
  \ket{\psi}=
  c_{00}\ket{00}+c_{01}\ket{01}+c_{10}\ket{10}+c_{11}\ket{11},
 \label{sol1}
\end{equation}
with the time-dependent probability amplitudes:
\begin{eqnarray}
c_{00} &=&\frac{1}{4} e^{-i (2 F+J) t} \left(1+e^{4 i F t}+2 e^{2 i (F+J) t}\right), \notag \\
c_{01} &=&c_{10}=-\frac{i}{2}  \exp(-i J t) \sin (2 F t), \notag \\
c_{11} &=&c_{00}-\exp(i J t), \label{sol1c}
\end{eqnarray}
as calculated for simplicity in the two-qubit Hilbert space. Only
for short evolution times, these solutions approximate well our
precise numerical solutions, which were obtained in a
high-dimensional Hilbert space and plotted in the left frames of
Fig.~\ref{fig07}. Much better agreement with these precise
numerical solutions can be found by diagonalizing the Hamiltonian
$H'_{\rm eff}$ in a two-qutrit Hilbert space. Unfortunately, we
cannot obtain a compact-form analytical solution in this case, as
discussed in Appendix~B.

As a measure of the quality of phonon blockade (or phonon
truncation), one can calculate the fidelity, defined as
$F(t)=\sum_{m_1,m_2=0,1}|c_{m_1,m_2}|^2$, where the probabilities
$|c_{m_1,m_2}|^2$ are computed precisely in a large-dimensional
Hilbert space. For the same parameters as in the left frames of
Fig.~\ref{fig07}, we find that the fidelity periodically
oscillates between the values 0.977 and 1. This shows that the
evolution of phonons in the NAMRs is practically confined in a
two-qubit Hilbert space in model~1 even without dissipation.

\begin{figure} 
\includegraphics[width=4.25cm]{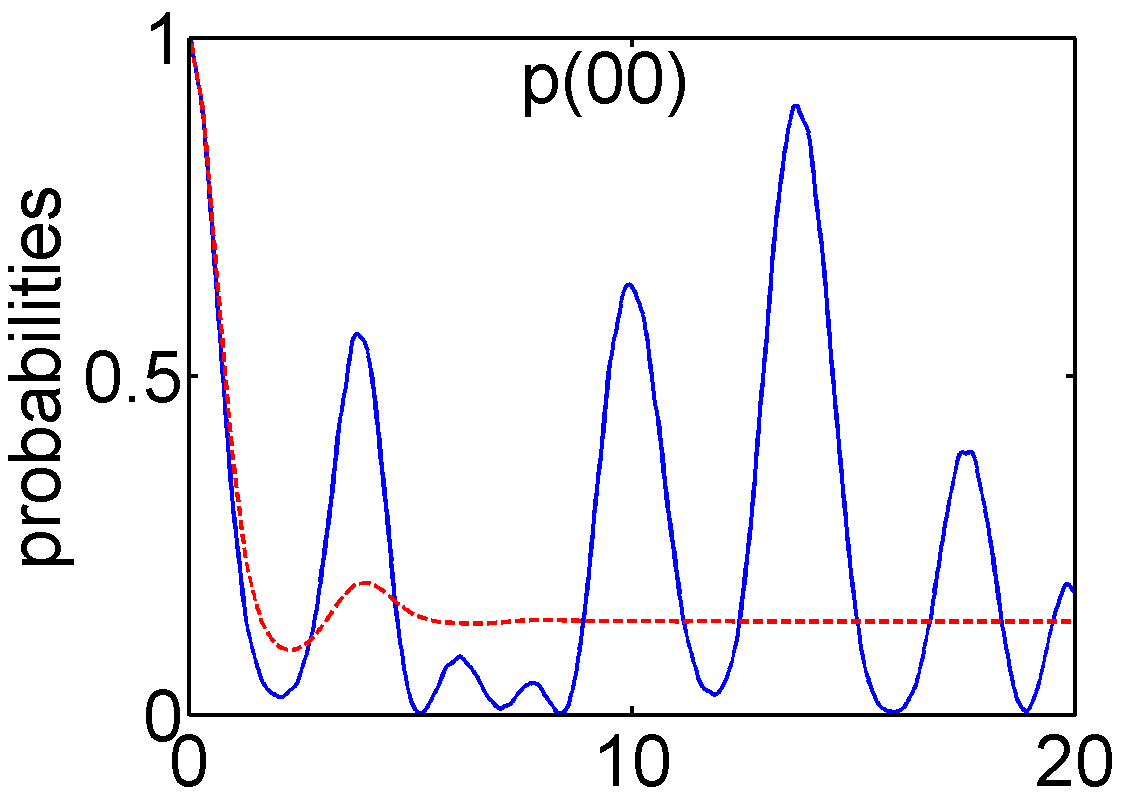}
\includegraphics[width=4cm]{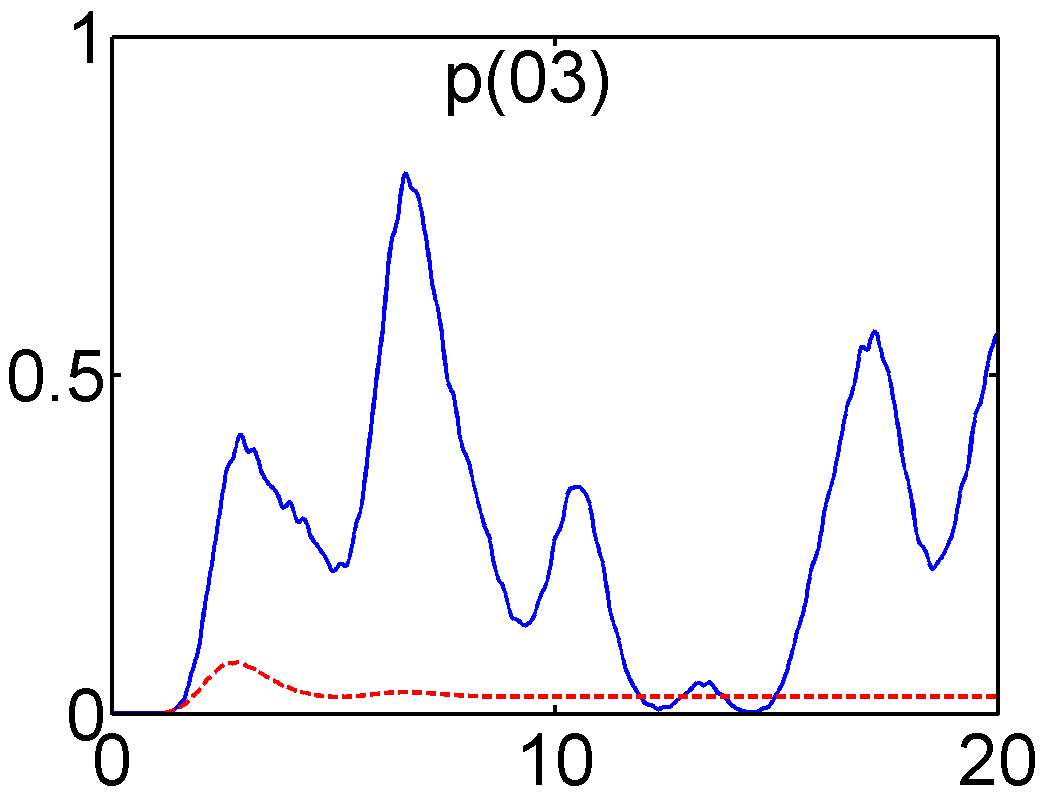}

\includegraphics[width=4.25cm]{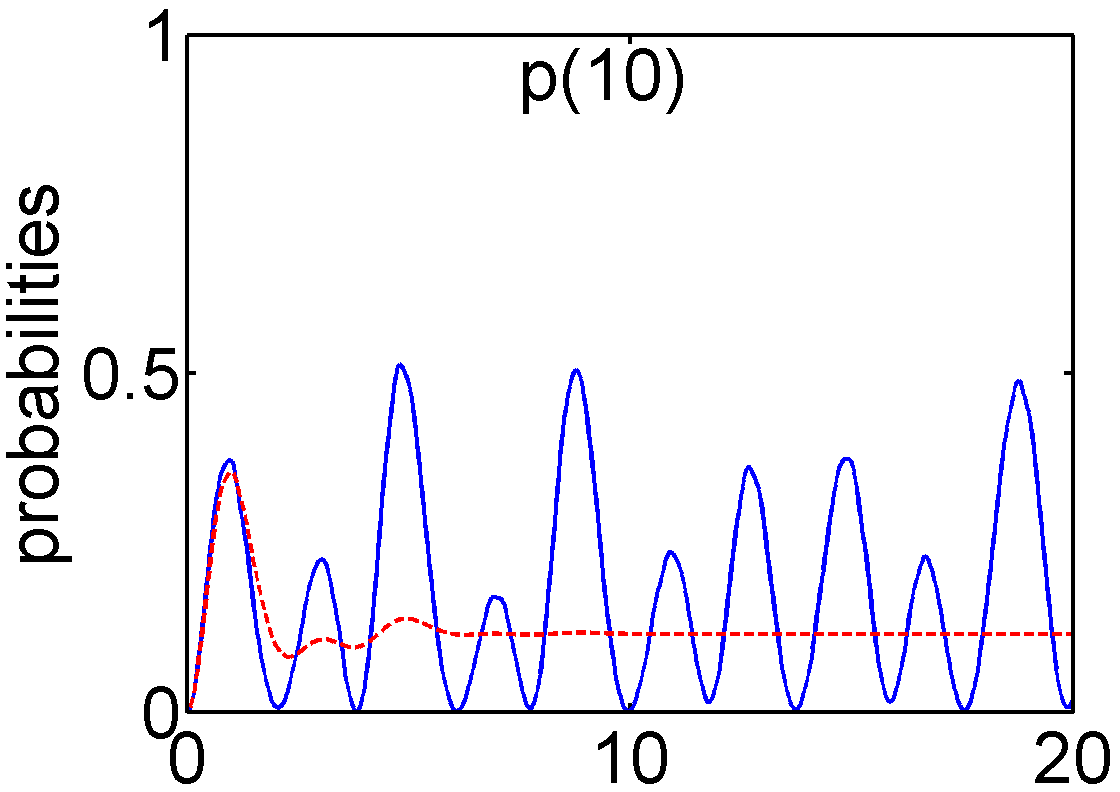}
\includegraphics[width=4cm]{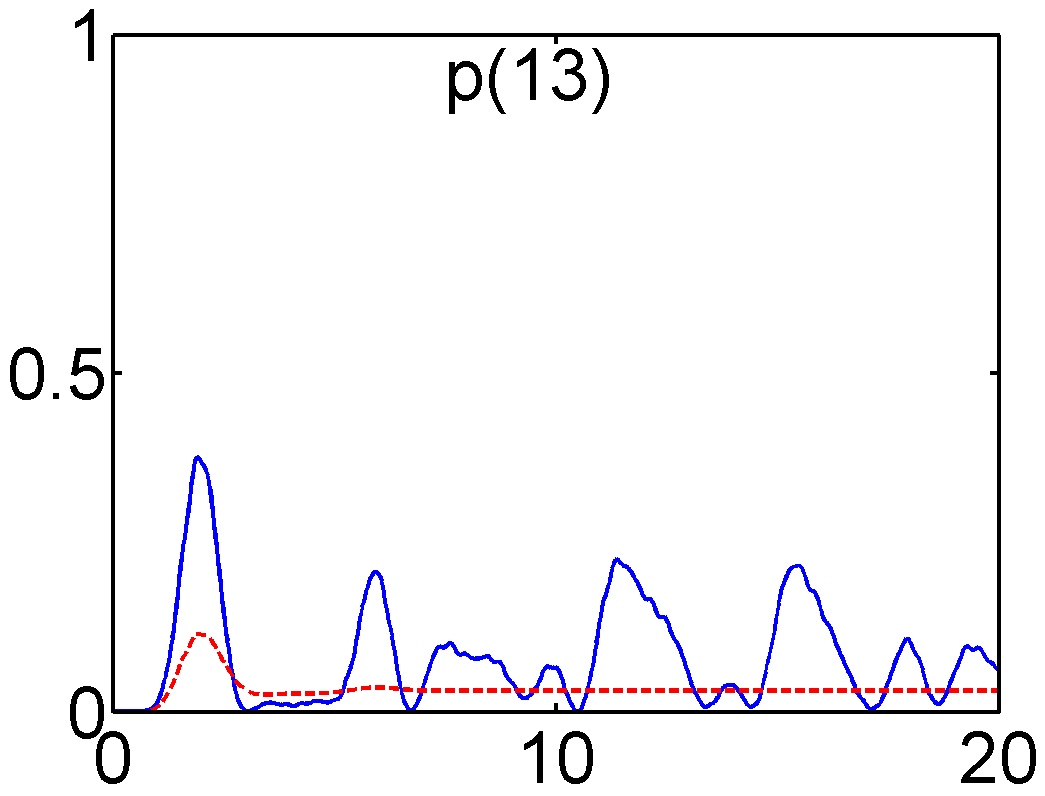}

\includegraphics[width=4.2cm]{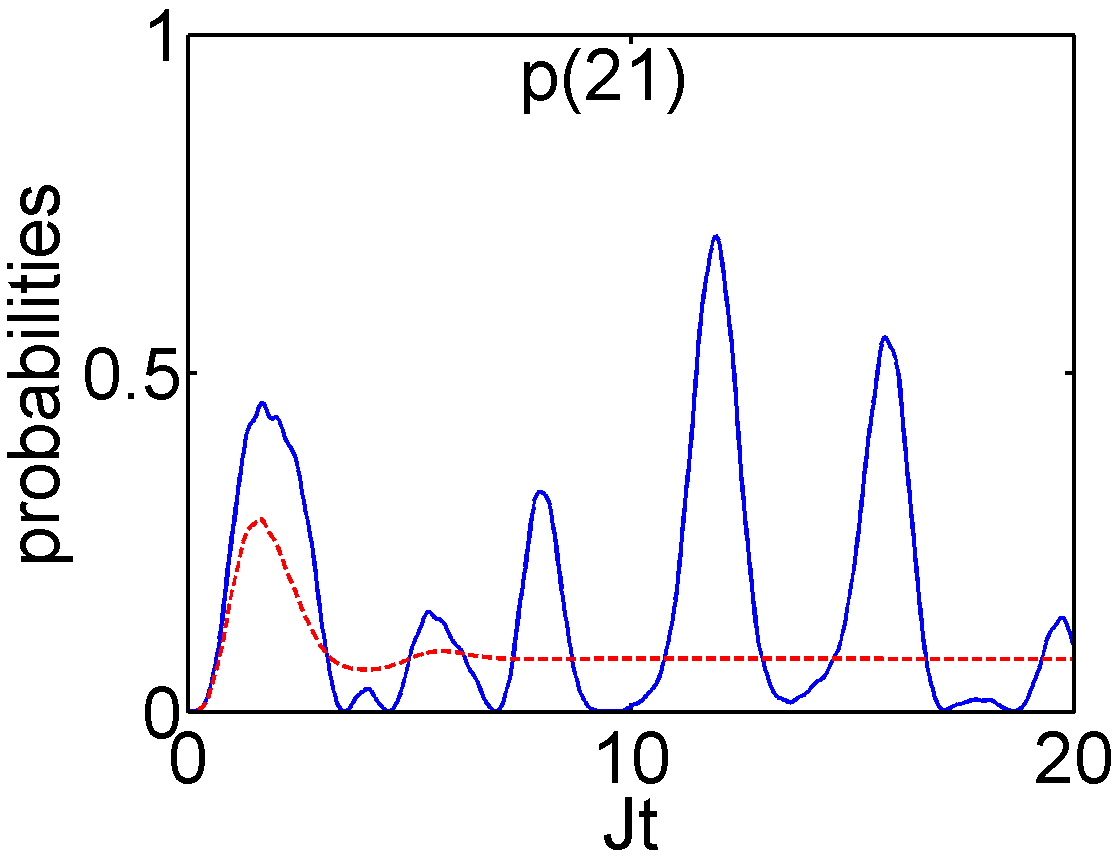}
\includegraphics[width=4cm]{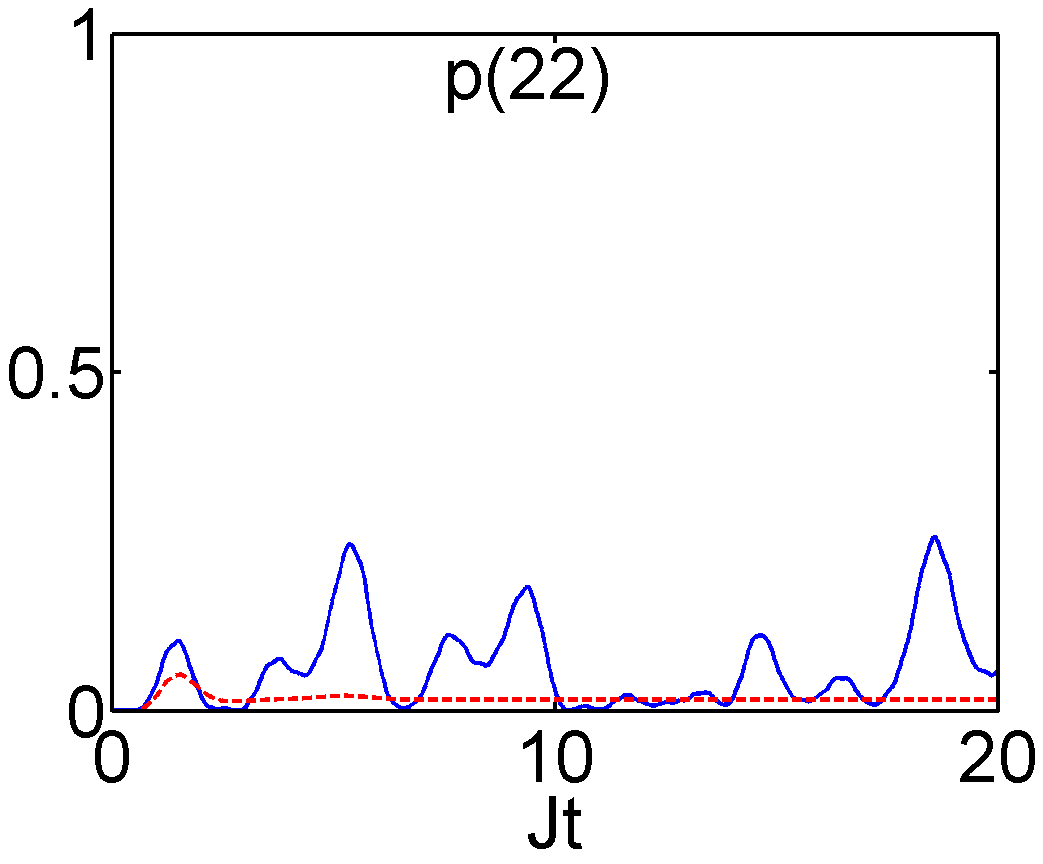}
\caption{(Color online) Non-stationary phonon blockade in model~2:
Probabilities $P(m_1,m_2)$ same as in Fig.~\ref{fig07} but for the
non-dissipative (blue solid curves, $\gamma=0$) and dissipative
(red dashed curves, $\gamma=J/3$) NAMR systems described by the
effective Hamiltonian $H''_{\rm eff}$. } \label{fig08}
\end{figure}

Let us now analyze phonon blockade in the non-dissipative model~2.
For simplicity, we again assume that the driving field strengths
are the same, $F_1=F_2\equiv F$. The solution of the Schr\"odinger
equation for the wave function $\ket{\psi}=\exp(-iH''_{\rm
eff}t)\ket{00}$ reads,
\begin{eqnarray}
  \ket{\psi}&=&
  c_{00}\ket{00}+c_{03}\ket{03}+c_{10}\ket{10}
  \notag \\
  && +\, c_{13}\ket{13}+c_{21}\ket{21}+c_{22}\ket{22},
 \label{sol2}
\end{eqnarray}
where the probability amplitudes can be found only numerically, as
explained in Appendix~B. We plotted the evolution of these
probability amplitudes for model~2 in Fig.~\ref{fig08} (solid blue
curves) in analogy to those shown in Fig.~\ref{fig07} for model~1.

Analogously to model~1, we can quantify the quality of phonon
blockade in model~2 by calculating the fidelity
$F(t)=\sum_{(m_1,m_2)}|c_{m_1,m_2}|^2$; but now
$(m_1,m_2)=(0,0),(0,3),(1,0),(1,3),(2,1),(2,2)$. For the
dissipation-free evolution shown in Fig.~\ref{fig08}, we find that
the fidelity $F(t)\in[0.9643,1]$. Thus, we can conclude that the
evolution of phonons in the NAMRs according to model~2 even
without damping is effectively confined in the Hilbert space of an
entangled qutrit-quartit system.

\subsection{Phonon blockade in dissipative systems}

In the standard description of dissipation under Markov's
approximation, the evolution of the reduced density operator
$\rho(t)$ is governed by the master equation,
\begin{eqnarray}
\dot\rho&=& -i[ {H}_{\rm eff}^{}
 ,{\rho}]
 \nonumber \\ &&
 +\sum_n\frac{\gamma_n}{2}\bar{n}^{(n)}_{\rm th}
 (2 a_{n}^{\dagger}{\rho}  a_{n} - a_{n}
 a_{n}^{\dagger}{\rho}  -{\rho}  a_{n}
 a_{n}^{\dagger})
 \nonumber \\ &&
+\sum_n\frac{\gamma_n}{2}(\bar{n}^{(n)}_{\rm th}+1)(2 a_{n}{\rho}
 a_{n}^{\dagger}
 - a_{n}^{\dagger} a_{n}{\rho}
 -{\rho}  a_{n}^{\dagger} a_{n}),\qquad
 \label{ME}
\end{eqnarray}
where $\gamma_n$ is the $n$th NAMR decay rate (damping constant),
$\bar{n}_{\rm th}^{(n)}=\{\exp[\omega/(k_{B}T)]-1\}^{-1}$ is the
mean number of thermal phonons interacting with the $n$th NAMR,
$k_{B}$ is the Boltzmann constant, and $T$ is the reservoir
temperature at thermal equilibrium. For simplicity, we assume
equal decay rates $\gamma_1=\gamma_2\equiv\gamma$ and mean
thermal-phonon numbers $\bar{n}_{\rm th}^{(1)}=\bar{n}_{\rm
th}^{(2)}\equiv \bar{n}_{\rm th}$. In our numerical analysis, we
focus on the steady-state solutions ${\rho}_{{\rm ss}}=
\rho(t\rightarrow \infty)$ of the master equation, obtained for
$\dot\rho=0$. We obtain such numerical solutions by applying the
inverse power method implemented in Ref.~\cite{SzeTan99}.

\begin{figure} 
\fig{\includegraphics[width=6cm]{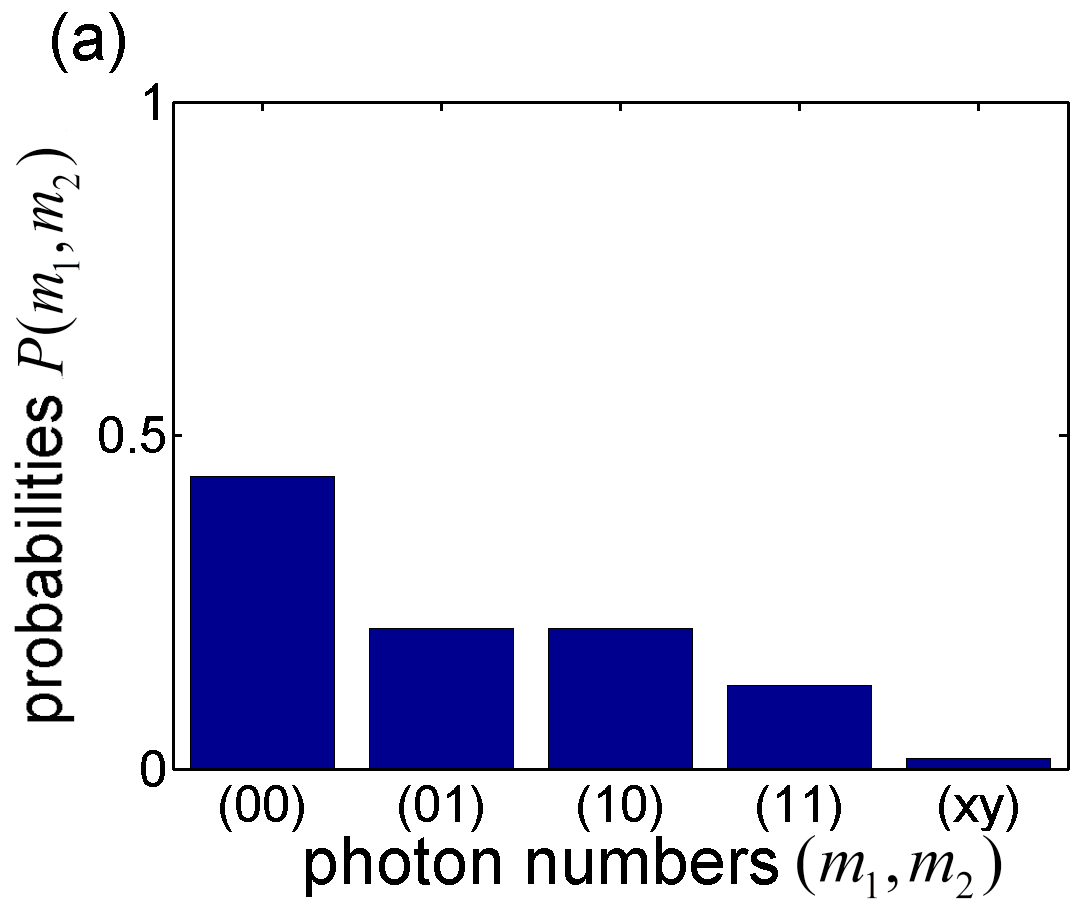}}

\fig{\includegraphics[width=6cm]{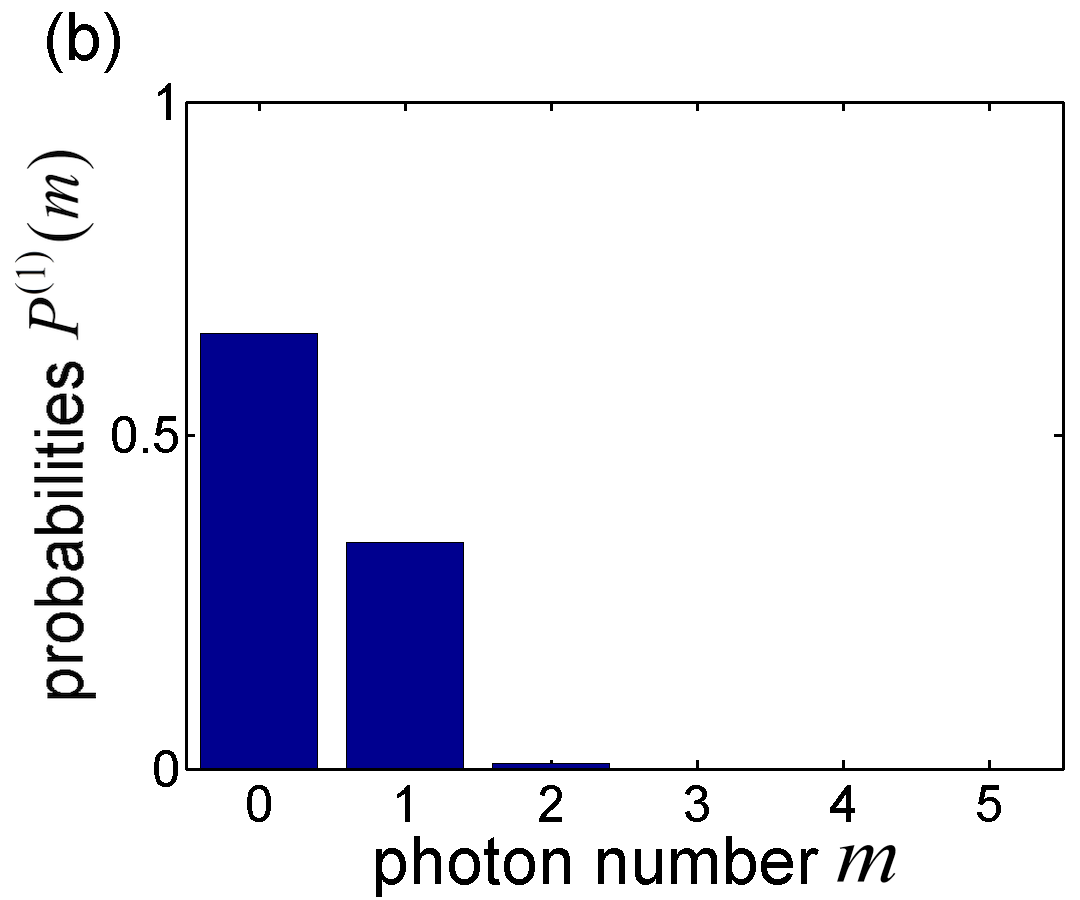}}

\caption{(Color online) Stationary solutions for model~1
describing an effective two-qubit system and, thus, corresponding
to single-phonon blockade: The phonon-number probabilities (a)
$P(m_1,m_2)=\bra{m_1,m_2}\rho_{\rm ss}\ket{m_1,m_2}$ for two NAMRs
and (b) $P^{(1)}(m)=\bra{m}{\rm Tr}_2(\rho_{\rm ss})\ket{m}$ for
the first NAMR (and, equivalently, $P^{(2)}(m)$ for the second
NAMR) for the steady-state solutions $\rho_{{\rm ss}}$ of the
master equation~(\ref{ME}) with the Hamiltonian $H_{\rm eff}'$,
given by Eq.~(\ref{Heff1}), assuming the same parameters as in
Fig.~\ref{fig07}. Moreover, $(xy)$ denotes all phonon numbers such
that $x,y>1$, so $P(xy)=1-\sum_{m_1,m_2=0,1}P(m_1,m_2)$. Panel (b)
clearly shows the occurrence of single-phonon blockade in every
NAMR.} \label{fig09}
\end{figure}

\begin{figure} 
\fig{\includegraphics[width=6cm]{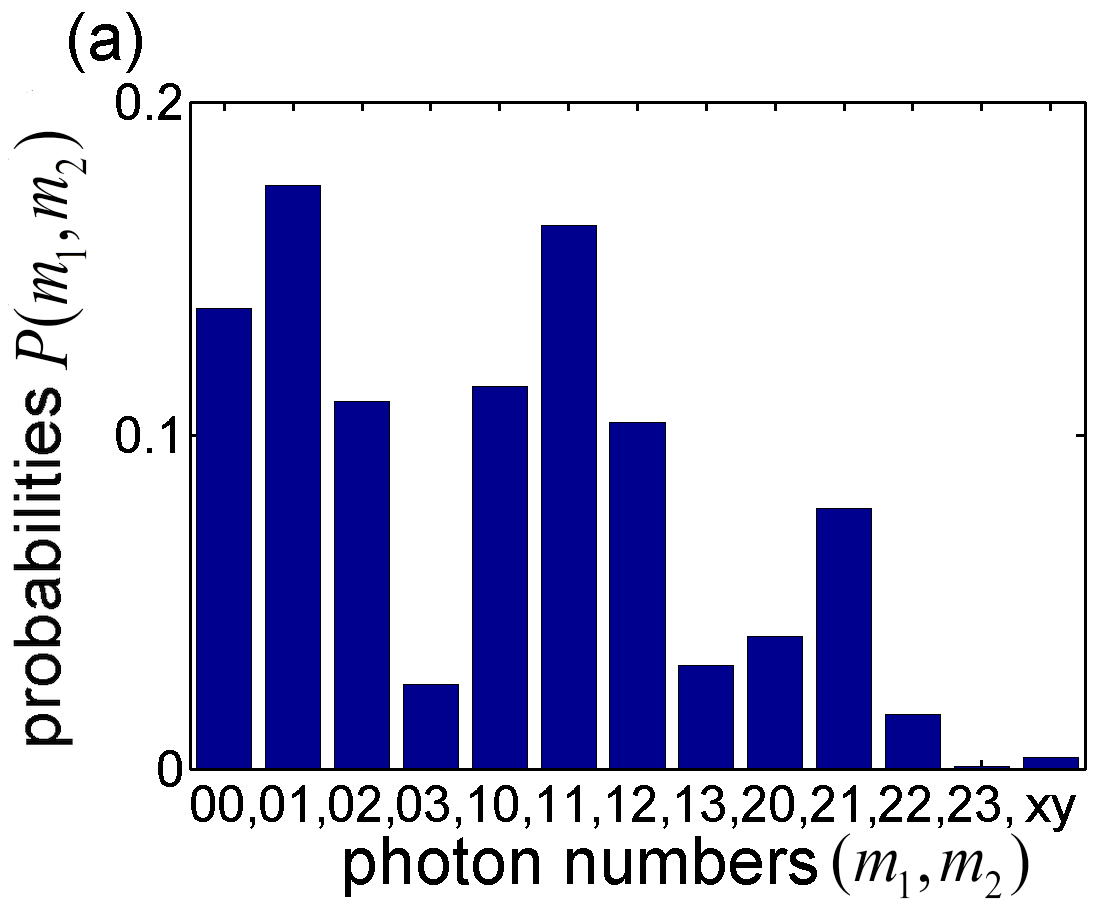}}

\fig{\includegraphics[width=6cm]{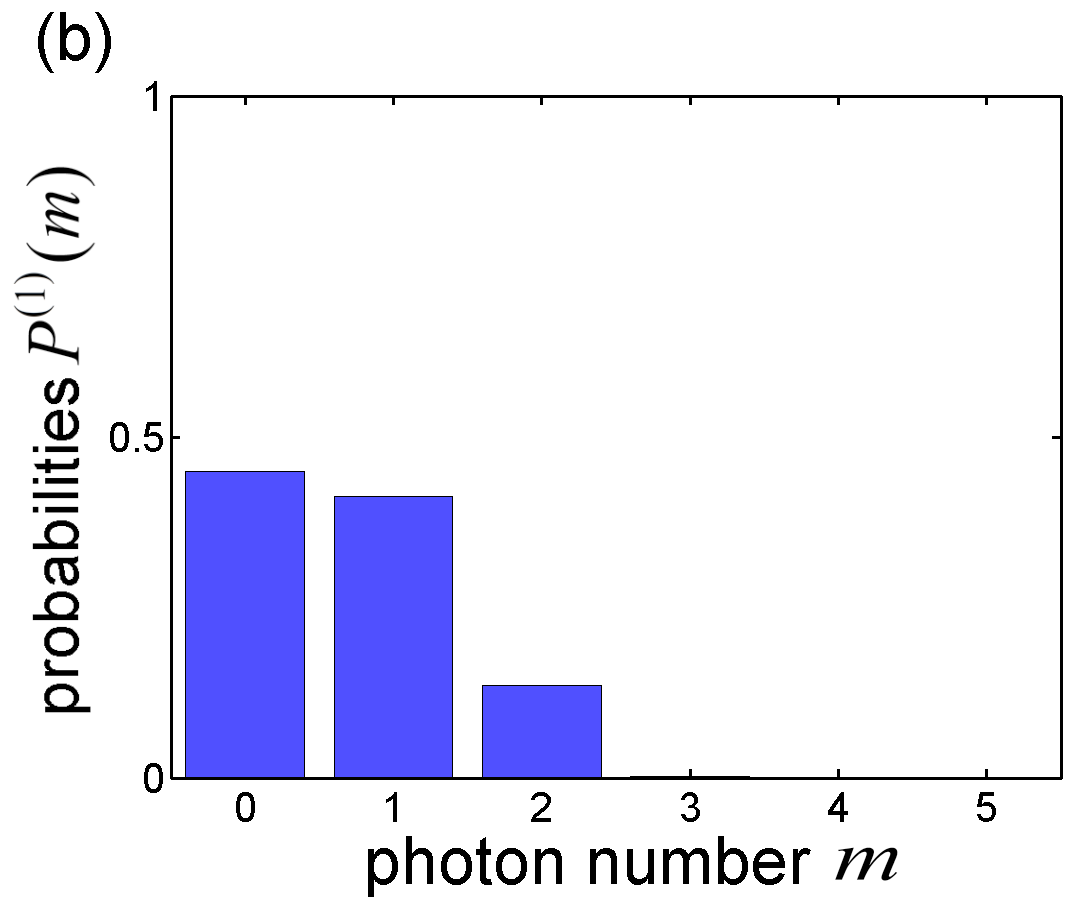}}

\fig{\includegraphics[width=6cm]{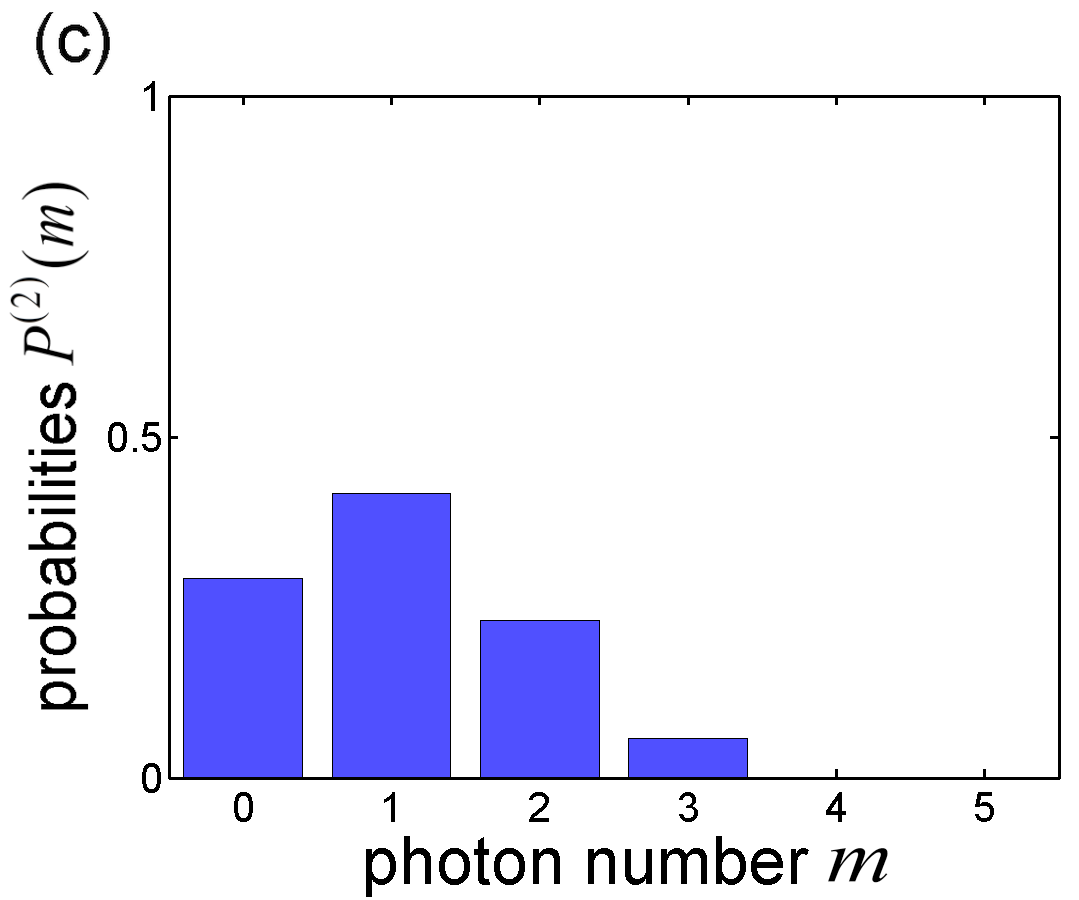}}

\caption{(Color online) Stationary solutions for model~2
describing an effective qutrit-quartit system and, thus,
corresponding to multiphonon blockade:  The phonon-number
probabilities (a) $P(m_1,m_2)=\bra{m_1,m_2}\rho_{\rm
ss}\ket{m_1,m_2}$ for two NAMRs, and (b,c) $P^{(n)}(m)=\bra{m}{\rm
Tr}_{3-n}(\rho_{\rm ss})\ket{m}$ for the $n$th NAMR ($n=1,2$) for
the steady-state solutions $\rho_{{\rm ss}}$ of the master
equation~(\ref{ME}) with the Hamiltonian $H_{\rm eff}''$, given by
Eq.~(\ref{Heff2}), and the same parameters as in Fig.~\ref{fig07}
and $(xy)$ denotes all phonon numbers such that $x>2$ and $y>3$.
Panel (b) [(c)] demonstrates the occurrence of two-phonon
(three-phonon) blockade in the first (second) NAMR.} \label{fig10}
\end{figure}

Examples of dissipative evolutions of phonon-number probabilities
are shown by the red curves in Fig.~\ref{fig07} for model~1 and
the red dashed curves in Fig.~\ref{fig08} for model~2. It is seen
that short-time oscillations are rapidly damped. The resulting
steady-state phonon-number probabilities are shown in
Figs.~\ref{fig09} and~\ref{fig10}, respectively. In panels (a) of
these figures, we plotted the probabilities
$P(m_1,m_2)=\bra{m_1,m_2}\rho_{\rm ss}\ket{m_1,m_2}$ for the
two-NAMR density matrices $\rho_{\rm ss}$. While in panels (b) and
(c), we presented the single-NAMR phonon-number probabilities
$P^{(n)}(m)=\bra{m}\rho^{(n)}_{\rm ss}\ket{m}$ for
$\rho^{(n)}_{\rm ss}={\rm Tr}_{3-n}(\rho_{\rm ss})$, with $n=1,2$.

We can clearly interpret these results as single-, two-, and
three-phonon blockades corresponding to the cases shown in
Figs.~\ref{fig09}(b), \ref{fig10}(b), and~\ref{fig10}(c),
respectively.

We note that the steady states are not pure, contrary to the
standard assumptions made in analogous studies of single-photon
blockades of optical~\cite{Bamba11,Xu14coupler} and
optomechanical~\cite{Wang15} systems. Indeed for the examples of
the states shown in Figs.~\ref{fig09} and~\ref{fig10}, we found
that their purities are the following: ${\rm Tr}\rho_{\rm
ss}^2=0.4212$ and ${\rm Tr}(\rho^{(1)}_{\rm ss})^2={\rm
Tr}(\rho^{(2)}_{\rm ss})^2=0.5670$ for model~1, and ${\rm
Tr}\rho_{\rm ss}^2=0.1471$, ${\rm Tr}(\rho^{(1)}_{\rm
ss})^2=0.3920$, and ${\rm Tr}(\rho^{(2)}_{\rm ss})^2=0.3212$ for
model~2.

Finally we note that, for simplicity, we applied here the standard
master equation, given by Eq.~(\ref{ME}) assuming two separable
dissipation channels for the NAMRs.  A more precise description,
which could be especially important for a stronger coupling
between the NAMRs, should be based on a generalized master
equation within the general formalism of Breuer and Petruzzione
(see sect. 3.3 in Ref.~\cite{BreuerBook}). In this approach both
NAMRs dissipate into usually-entangled dissipation channels. An
explicit form of such generalized master equation for two
strongly-coupled infinitely-dimensional systems will be presented
elsewhere~\cite{Miran16}. Note that a generalized master equation
for an infinitely-dimensional system strongly coupled to a qubit
system has already been well studied~\cite{Beaudoin11}.


\section{Phonon blockade in phase space}

Here we apply the Wigner function $W\equiv W^{(0)}$ and the
Cahill-Glauber $s$-parametrized quasiprobability distribution
(QPD) $W^{(s)}$ in order to visualize the nonclassical properties
of the phonon-blockaded states studied in Sec.~III.

The Wigner function for a two-mode (or two-NAMR) state $\rho$ can
be given by
\begin{eqnarray}
W_{12}(\alpha_1,\alpha_2)&=& W_{12}(q_1,p_1,q_2,p_2)
\notag \\
&=&\frac{1}{\pi^2}\int\bra{q_1-x_1,q_2-x_2}\rho
\ket{q_1+x_1,q_2+x_2}
\notag \\
&&\times\exp\left[2i(p_1x_1+p_2x_2) \right]dx_1dx_2,
\label{Wigner2}
\end{eqnarray}
in terms of the canonical position $q_n$ and momentum $p_n$
operators, and $\alpha_n=q_n+ip_n$ for the $n$th NAMR. It is seen
that Eq.~(\ref{Wigner2}) is a straightforward generalization of
the Wigner function for a single-mode (in our case single-NAMR)
case,
\begin{eqnarray}
W(\alpha_n)= W(q_n,p_n)=\frac{1}{\pi}\int\bra{q_n-x_n}\rho_n
\ket{q_n+x_n}\notag \\ \times\exp\left(2ip_nx_n\right)dx_n,
\label{Wigner1}
\end{eqnarray}
where $\rho_n\equiv \rho^{(n)}$ can correspond, e.g., to
$\tr_{3-n}\rho$ for $n=1,2$. Specifically, the single-NAMR Wigner
function $W(\alpha_n)$ can be considered as the marginal functions
of the two-NAMR Wigner function $W(\alpha_1,\alpha_2)$.

\begin{figure} 
\fig{\includegraphics[height=4.4cm]{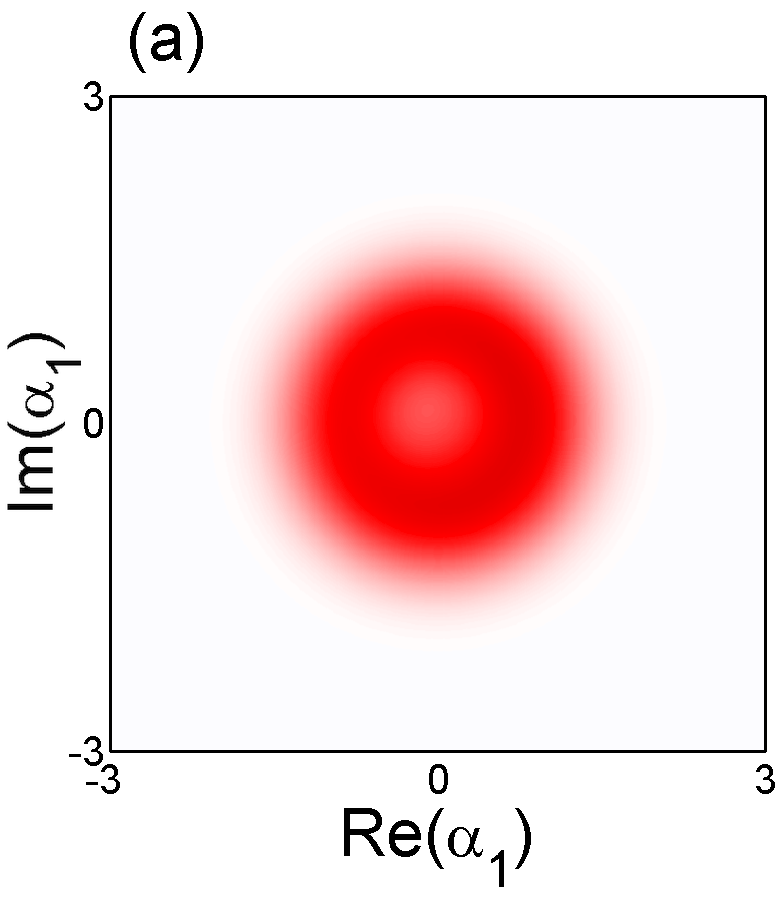}}
\fig{\includegraphics[height=4.4cm]{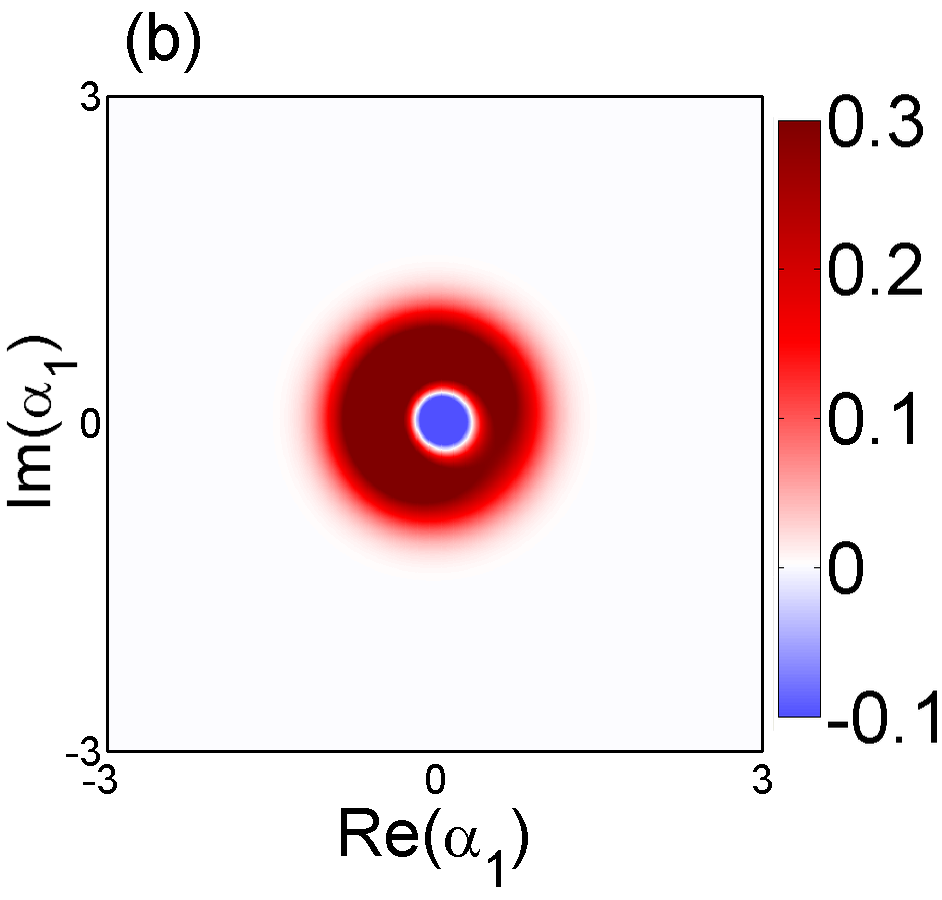}}

\caption{(Color online) Stationary solutions for model~1 as in
Fig.~\ref{fig09} but for (a) the single-NAMR Wigner function
$W(\alpha_1)=W(\alpha_2)$ and (b) the single-NAMR quasiprobability
distribution (QPD) function $W^{(s)}(\alpha_1)=W^{(s)}(\alpha_2)$
with parameter $s=1/2$ for the steady-state solutions
$\rho^{(n)}_{{\rm ss}}={\rm Tr}_{3-n}\rho_{{\rm ss}}$ (for
$n=1,2$). Note that the negative regions of the QPD functions are
marked in blue.} \label{fig11}
\end{figure}

Figure~\ref{fig11} shows the Wigner function
$W(\alpha_1)=W(\alpha_2)$ for the steady state $\rho^{(n)}_{{\rm
ss}}={\rm Tr}_{3-n}\rho_{{\rm ss}}$ ($n=1,2$) for some chosen
values of the coupling and damping parameters. Unfortunately, the
contribution of the vacuum state in $\rho_{{\rm ss}}$ is dominant,
as shown in Fig.~\ref{fig09}. Thus, the Wigner function for
$\rho^{(n)}_{{\rm ss}}$ looks like a slightly deformed Gaussian
representing the vacuum.

To show this deformation more clearly we also plotted the
$s$-parametrized Cahill-Glauber QPD, $W^{(s)}(\alpha_n)$ for
$s=1/2$. For simplicity, we analyze this QPD only for a
single-mode (i.e., single-NAMR) case, while the extension for the
two- and multi-mode cases is straightforward. The Cahill-Glauber
QPD $W^{(s)}(\alpha_n)$ can be defined in the Fock-state
representation of an arbitrary-dimensional single-mode state
$\rho$ as follows~\cite{Cahill69}:
\begin{equation}
  W^{(s)}(\alpha_n)=\sum_{k,l=0}^\infty\bra{k}\rho\ket{l}\bra{l}T^{(s)}(\alpha_n)\ket{k},
  \label{QPD1}
\end{equation}
where
\begin{equation}
  \bra{l}T^{(s)}(\alpha_n)\ket{k} = c
  \sqrt{\frac{l!}{k!}}y^{k-l+1}z^l(\alpha_n^*)^{k-l}L^{k-l}_l(x_{\alpha_n}),
\label{QPD2}
\end{equation}
for $s\in[-1,1]$, $c=\frac{1}{\pi}\exp[-2|\alpha_n|^2/(1-s)]$,
$x_{\alpha_n}=4|\alpha_n|^2/(1-s^2)$, $y=2/(1-s)$,
$z=(s+1)/(s-1)$, and $L^{k-l}_l$ are the associate Laguerre
polynomials. As for the Wigner function, $\alpha_n$ is a complex
number, where its real and imaginary parts can be interpreted as
canonical position and momentum, respectively. The operator
$T^{(s)}(\alpha_n)$ is defined in the Fock representation by
Eq.~(\ref{QPD2}). In the special cases of $s=-1,0,1$, the QPD
$W^{(s)}(\alpha_n)$ becomes the Husimi $Q$, Wigner $W$, and
Glauber-Sudarshan $P$~functions, respectively.

Here, we apply a well known definition of nonclassicality (see,
e.g., Ref.~\cite{Miran15} and references therein): A photonic or
phononic state can be considered nonclassical if and only if its
Glauber-Sudarshan $P$ function is not positive (semi)definite,
which means that it is not a classical probability density. Thus
only coherent states and their statistical mixtures are classical.

\begin{figure} 
\centerline{\includegraphics[height=4.4cm]{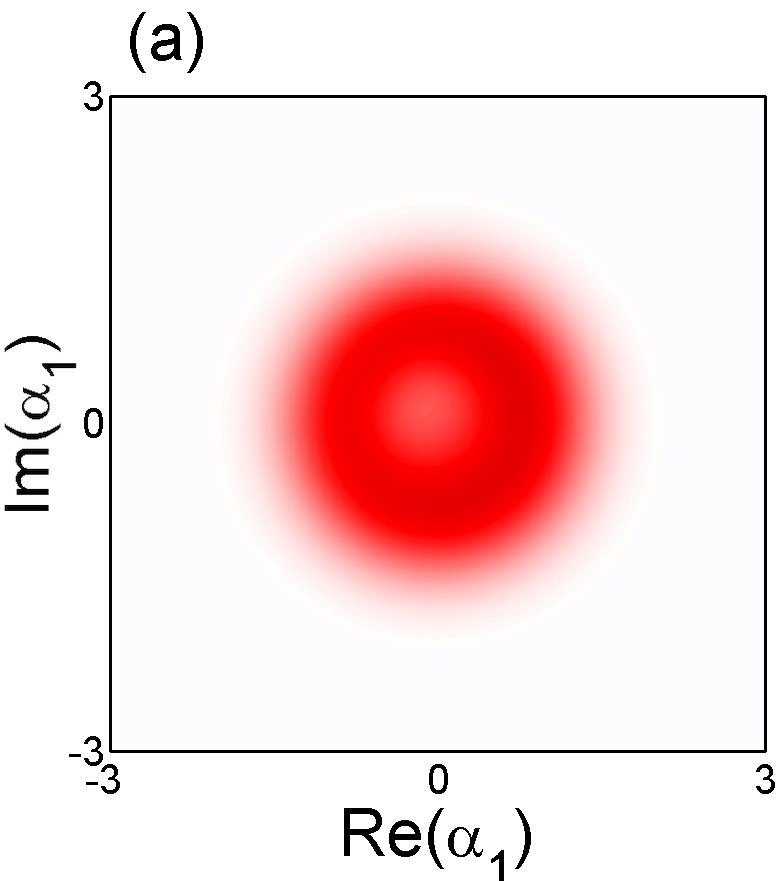}
\includegraphics[height=4.4cm]{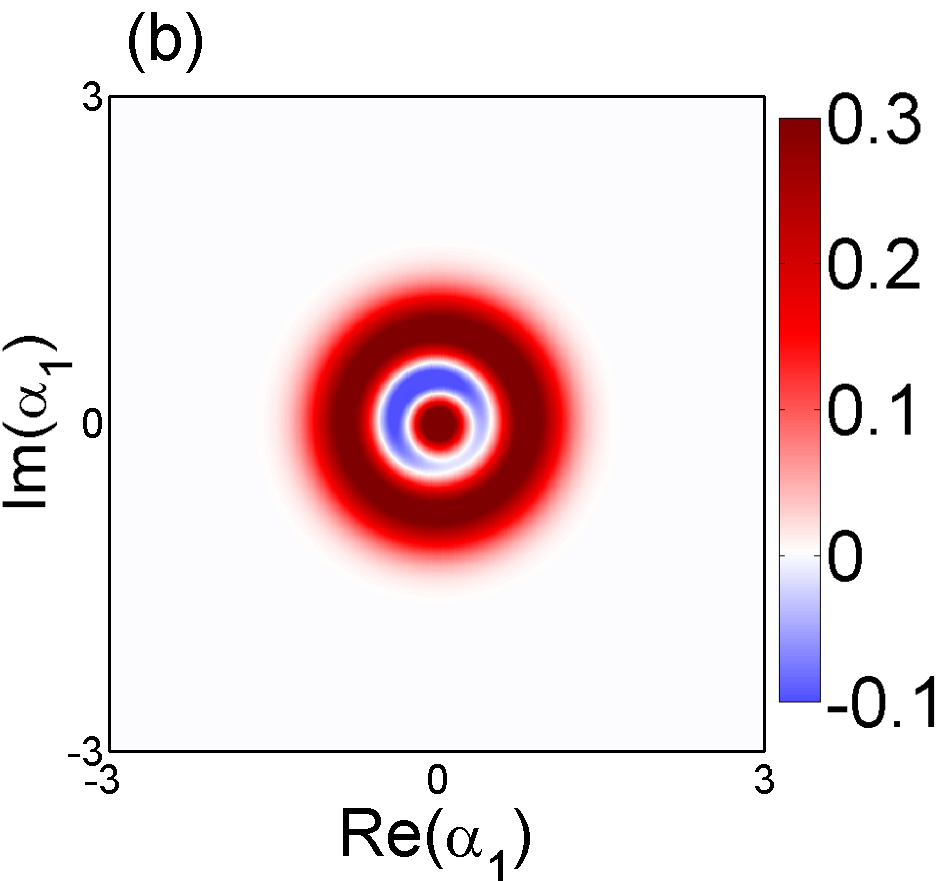}}

\centerline{\includegraphics[height=4.4cm]{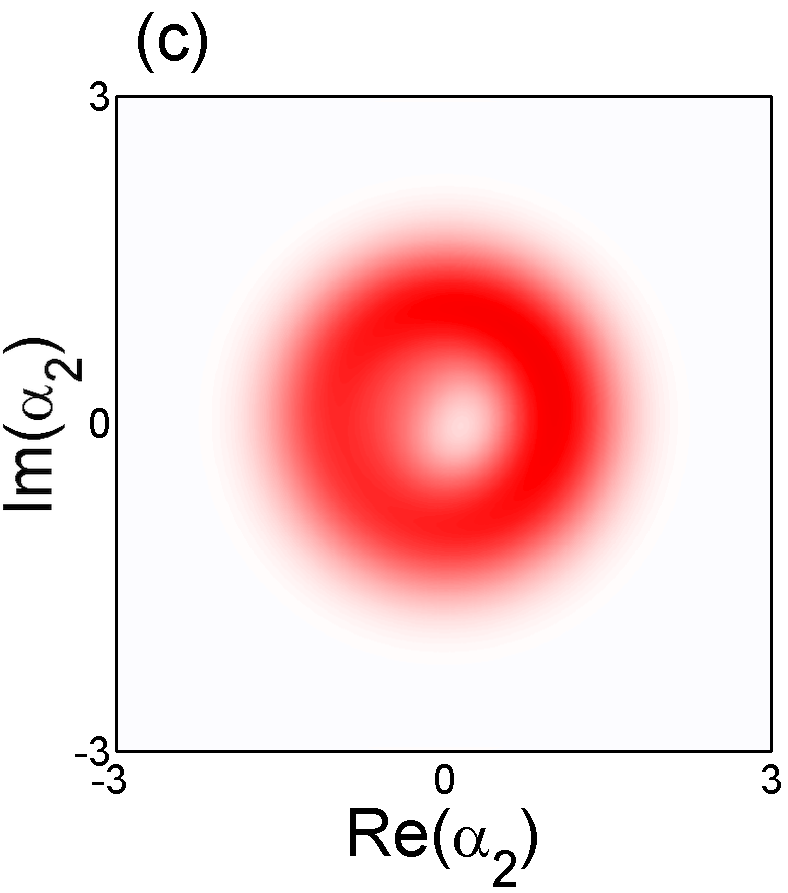}
\includegraphics[height=4.4cm]{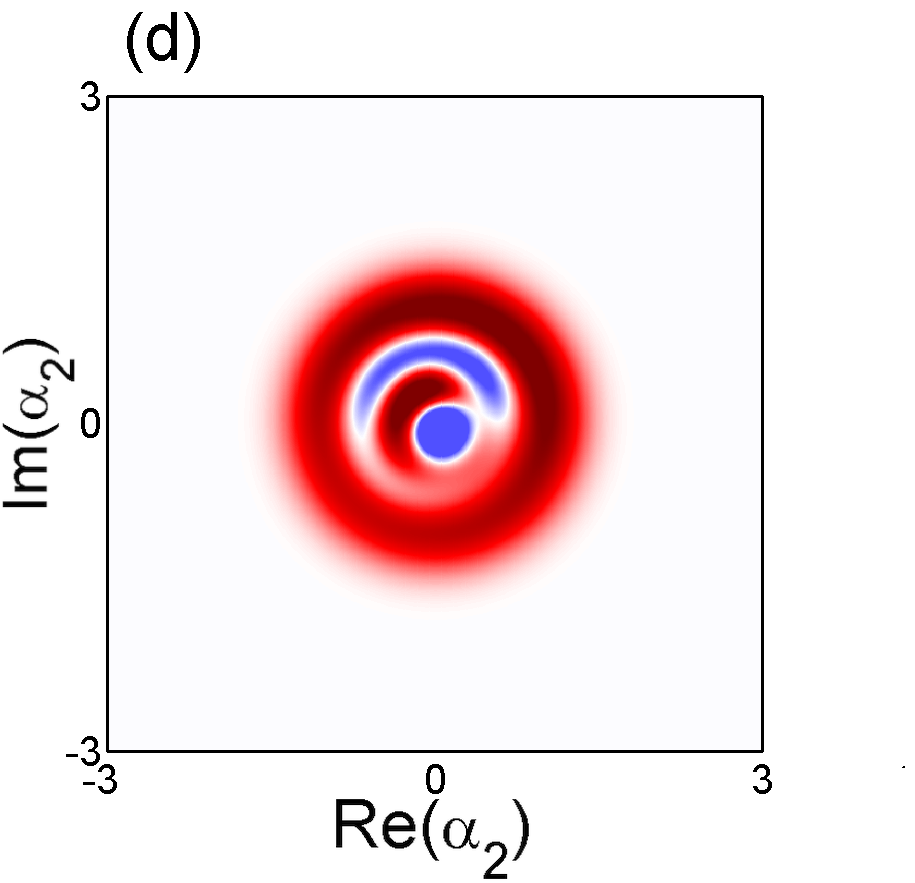}}

\caption{(Color online) Stationary solutions for model~2 as in
Fig.~\ref{fig10}, but for the single-NAMR Wigner functions
$W(\alpha_n)$ (a,c) and the single-NAMR QPD functions
$W^{(s)}(\alpha_n)$ with parameter s=1/2 (b,d) for the
steady-state solutions $\rho^{(n)}_{{\rm ss}}={\rm
Tr}_{3-n}\rho_{{\rm ss}}$ (for $n=1,2$). } \label{fig12}
\end{figure}

The Wigner functions and $1/2$-parametrized QDPs for the steady
states of models~1 and 2 are shown in Figs.~\ref{fig11}
and~\ref{fig12}, respectively. These steady states $\rho_{\rm ss}$
are nonclassical, as they correspond to the partially-incoherent
finite superpositions of phonon Fock states, which are not
mixtures of coherent states (in particular, they are not the
vacuum). The nonclassicality of these states $\rho_{\rm ss}$ is
clearly seen in the non-positive functions $W^{(s=1/2)}$ (their
negative regions are plotted in blue). However, the
nonclassicality of $\rho_{\rm ss}$ is difficult to deduce from the
plots of the nonnegative Wigner functions.

To understand this apparent discrepancy, we recall a known
relation between two single-mode QPDs,  ${\cal W}^{(s_0)}$ and
${\cal W}^{(s)}$ for the chosen parameters
$s<s_0$~\cite{Cahill69}:
\begin{equation}
{\cal W}^{(s)} ( \alpha_n) =c'\int \exp\left( - \frac{2| \alpha_n
- \beta_n|^2}{s_0-s} \right) {\cal W}^{(s_0)} ( \beta_n) {\rm d}^2
\beta_n, \label{QPD4}
\end{equation}
where $c'=2/[\pi(s_0-s)]$. This relation means that any QPD for
$s\le s_0$ can be obtained from ${\cal W}^{(s_0)}$ by mixing it
with the Gaussian noise. In particular, the Wigner function can be
obtained in this way from the $P={\cal W}^{(1)}$ and ${\cal
W}^{(1/2)}$ QPDs. By decreasing the parameter $s$ from $s_0=1$,
the QPD ${\cal W}^{(s)}$, for a given nonclassical state, becomes
less and less negative, and finally becomes nonnegative at some
$s' \ge -1$. As a result, in the analyzed examples shown in
Figs.~\ref{fig11} and \ref{fig12}, the negativity of the
$P$-function for $\rho_{\rm ss}$ is only partially lost in the
QPDs ${\cal W}^{(1/2)}$, but completely lost in the corresponding
Wigner functions.


\section{Entanglement, dimensionality, and nonclassicality of NAMRs}

To analyze more deeply the nonclassical properties of the
generated phonon states in the two NAMRs in models~1 and~2, we
apply the following measures of quantum correlations: the
negativity and its closely related entanglement dimensionality, as
well as the entanglement potential, as a measure of
nonclassicality.

\subsection{Entanglement}

To quantify the entanglement of a bipartite state $\rho$ of
arbitrary finite dimensions, we apply the negativity $N$, which
can be expressed as~\cite{Horodecki09review}
\begin{equation} \label{neg}    
 N(\rho) = \frac{\vert\vert\rho^{{\Gamma}}\vert\vert_{1}-1}{2}
\end{equation}
via the trace norm $\vert\vert\rho^{{\Gamma}}\vert\vert_{1}$ of
the partially-transposed statistical operator $ \rho^{{\Gamma}} $.
This entanglement measure is closely related to the
Peres-Horodecki criterion. The negativity $ N$ is an entanglement
monotone and, thus, can be used in quantifying entanglement in
bipartite systems. However, the negativity does not detect bound
entanglement (i.e., nondistillable entanglement) in systems more
complicated than two qubits or
qubit-qutrit~\cite{Horodecki09review}. The negativity can be
interpreted operationally. For example, the logarithmic
negativity,
\begin{equation}\label{Ecost}
 E_{\rm cost}(\rho) = \log_2[N(\rho) + 1],
\end{equation}
quantifies the entanglement cost under operations preserving the
positivity of the partial transpose (PPT), which is, for short,
referred to as the PPT entanglement
cost~\cite{Audenaert03,Ishizaka04}.

The evolutions of this entanglement measure are plotted in
Fig.~\ref{fig13}(a) for model~1 and Figs.~\ref{fig14}(a) for
model~2, by including and excluding the dissipation. The
oscillations of $E_{\rm cost}(\rho)$ are rapidly damped; however
the entanglement is not completely lost in the infinite-time
limit. Indeed, for the coupling parameters $K,J,F$, decay rate
$\gamma$, and thermal-phonon mean numbers $\bar{n}_{\rm th}$
specified in the figures, the entanglement between the NAMRs is
found to be $E_{\rm cost}(\rho_{\rm ss})=0.1413$ for model~1 and
almost three times smaller $E_{\rm cost}(\rho_{\rm ss})=0.0494$
for model~2.

\subsection{Dimensionality}

The negativity also determines the dimensionality $D_{\rm ent}$ of
entanglement, which is the number of degrees of freedom of two
entangled subsystems. Specifically, the entanglement
dimensionality $D_{\rm ent} $ for a bipartite state $\rho$ is
simply related to the negativity $N(\rho)$ as
follows~\cite{Eltschka13}
\begin{equation}
 D_{\rm ent}(\rho) = 2N(\rho) + 1=\vert\vert\rho^{{\Gamma}}\vert\vert_{1}.
 \label{Dent}
\end{equation}
More precisely, the least integer $\ge D_{\rm ent}$ gives a lower
bound to the number of entangled dimensions between the entangled
subsystems of $\rho$~\cite{Eltschka13}. According to
Eq.~(\ref{Dent}), $ D_{\rm ent}=1$ for separable states ($ N=0 $).
This measure could be useful for characterizing even a single test
system (in our case, a single phonon mode) with unknown quantum
dimension. This can be done in a standard way ``by entangling [the
test system] with an auxiliary system of known dimension and
measuring the negativity, a lower bound to the number of quantum
levels in the test system can be found''~\cite{Eltschka13}. In our
case of two NAMRs, we can directly apply the negativity, without
the use of an auxiliary system, to determine a lower bound to the
number of quantum levels in the total system (see also
Ref.~\cite{Arkhipov15}).

The evolutions of the entanglement dimensionality are plotted in
Fig.~\ref{fig13}(b) for model~1 and Fig.~\ref{fig14}(b) for
model~2. Since the entanglement dimensionality and the
entanglement cost are closely related, we can conclude, the same
as for $E_{\rm cost}(\rho)$,  that $ D_{\rm ent}$ does vanish in
the steady states. Specifically, the entanglement dimensionality
between the NAMRs reads: $D_{\rm ent}(\rho_{\rm ss})=1.2058$ for
model~1 and $D_{\rm ent}(\rho_{\rm ss})=1.0696$ for model~2.

\subsection{Nonclassicality}

The negativity can also be used in quantifying the nonclassicality
of a single-mode photonic or phononic state $\rho_n$ via the
so-called entanglement potential (EP), which is defined
as~\cite{Asboth05,Miran15}
\begin{equation}
  \EP(\rho_n) \equiv \log_2\big\{ N\big[\exp(-iHt)
  (\rho_n\otimes \ket{0}\bra{0} )\exp(iHt)\big]+1\big\}. \label{NP}
\end{equation}
Here  $H=\frac12(a_n^\dagger b+a_n b^\dagger)$ describes a
balanced beam splitter or a linear coupler, where $a_n$ and $b$
are the annihilation operators of the input modes. The basic idea
behind this measure in optics is as follows: If a single-mode
nonclassical (classical) photonic state is combined with the
vacuum at a beam splitter then the output state is entangled
(separable), for which various entangled measures (including the
negativity) can be applied. By generalizing this concept for
phonons it is enough to interpret this ancilla beam splitter as a
linear coupler.

\begin{figure} 
\includegraphics[width=6cm]{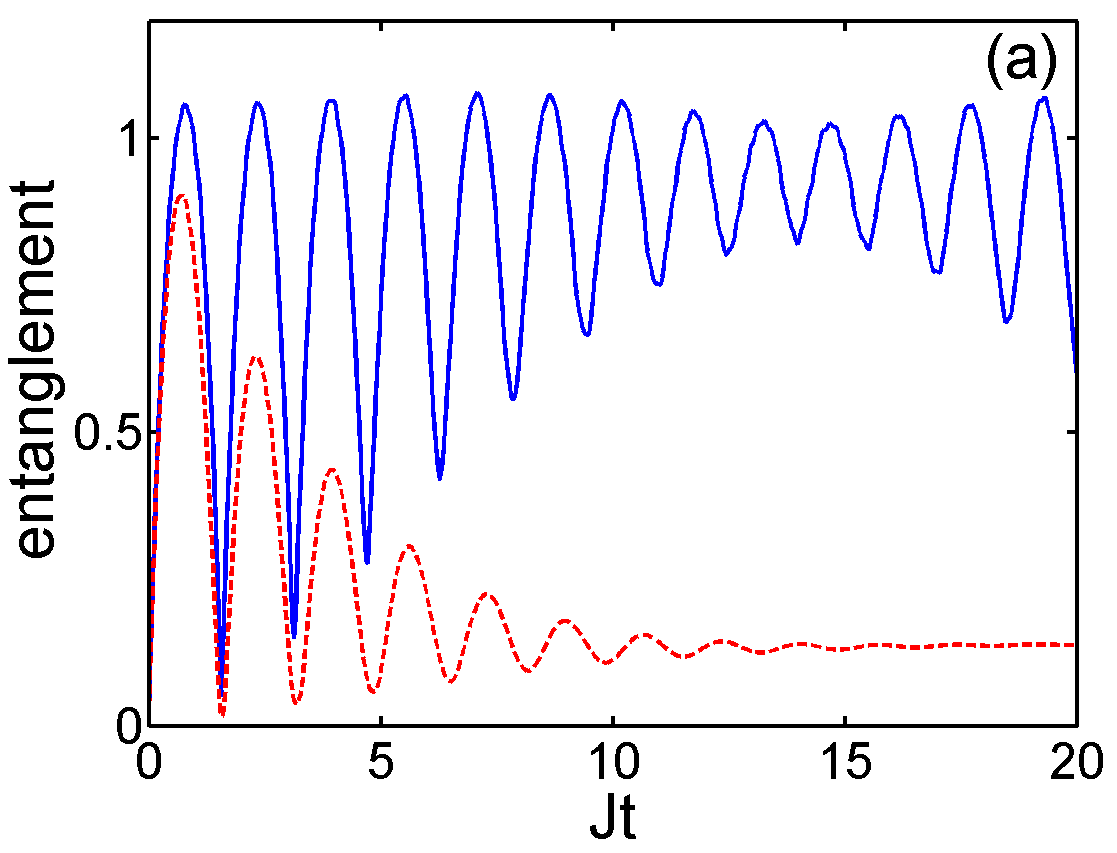}

\includegraphics[width=6cm]{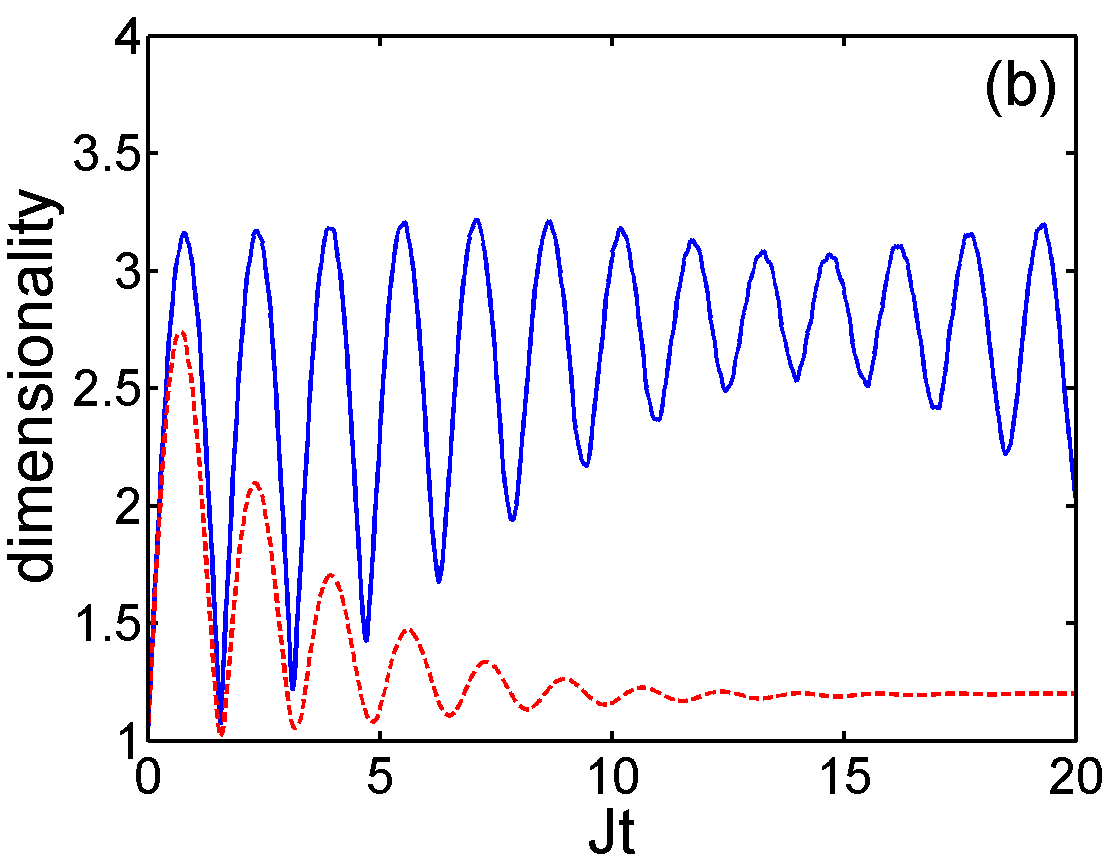}

\includegraphics[width=6cm]{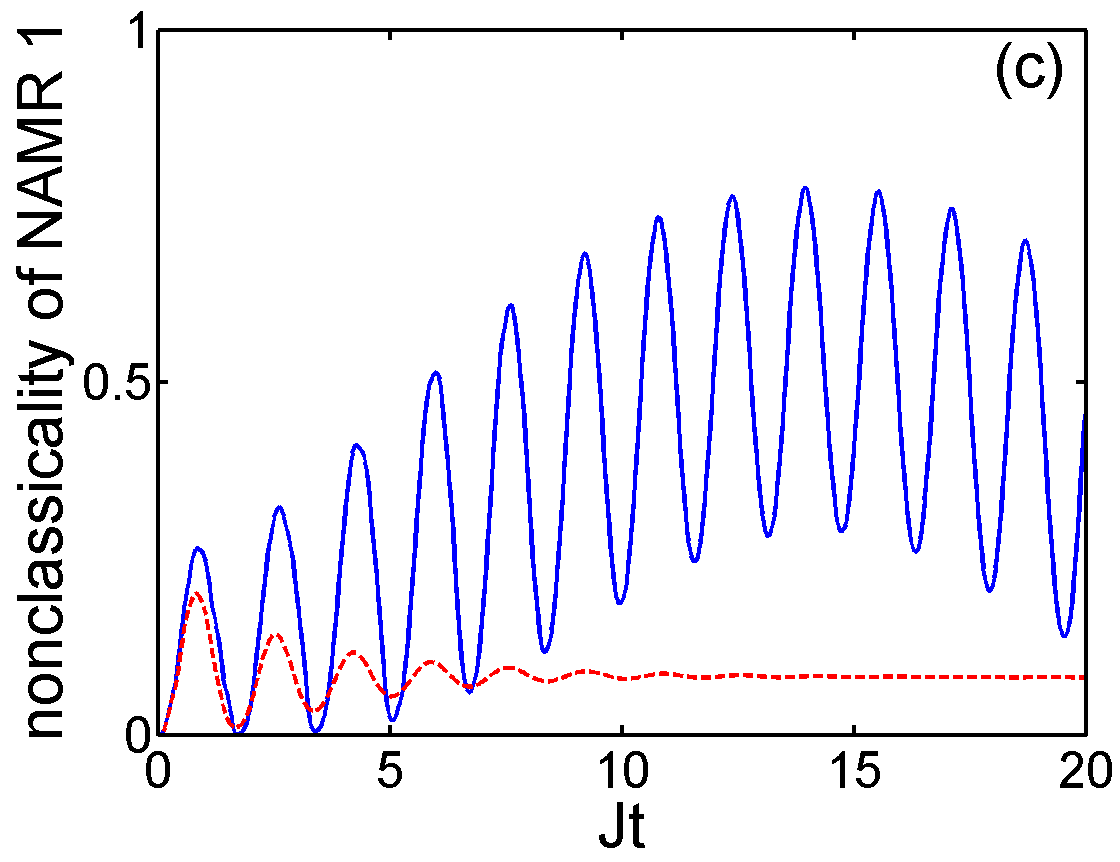}
\caption{(Color online) Non-stationary solutions for model~1: (a)
Entanglement, measured by the PPT entanglement cost $E_{\rm
cost}(\rho)$, (b) dimensionality of entanglement $D_{\rm
dim}(\rho)$, and (c) nonclassicality, measured by the entanglement
potential $\EP(\rho^{(1)})$, of the first (and, equivalently,
second) NAMR  for the states $\rho$ generated in the
non-dissipative (blue solid upper curves, $\gamma=0$) and
dissipative (red dashed lower curves, $\gamma=J/3$) systems
described by the effective Hamiltonian $H'_{\rm eff}$. Parameters
are same as in Fig.~\ref{fig07}. } \label{fig13}
\end{figure}

\begin{figure} 
\includegraphics[width=6.2cm]{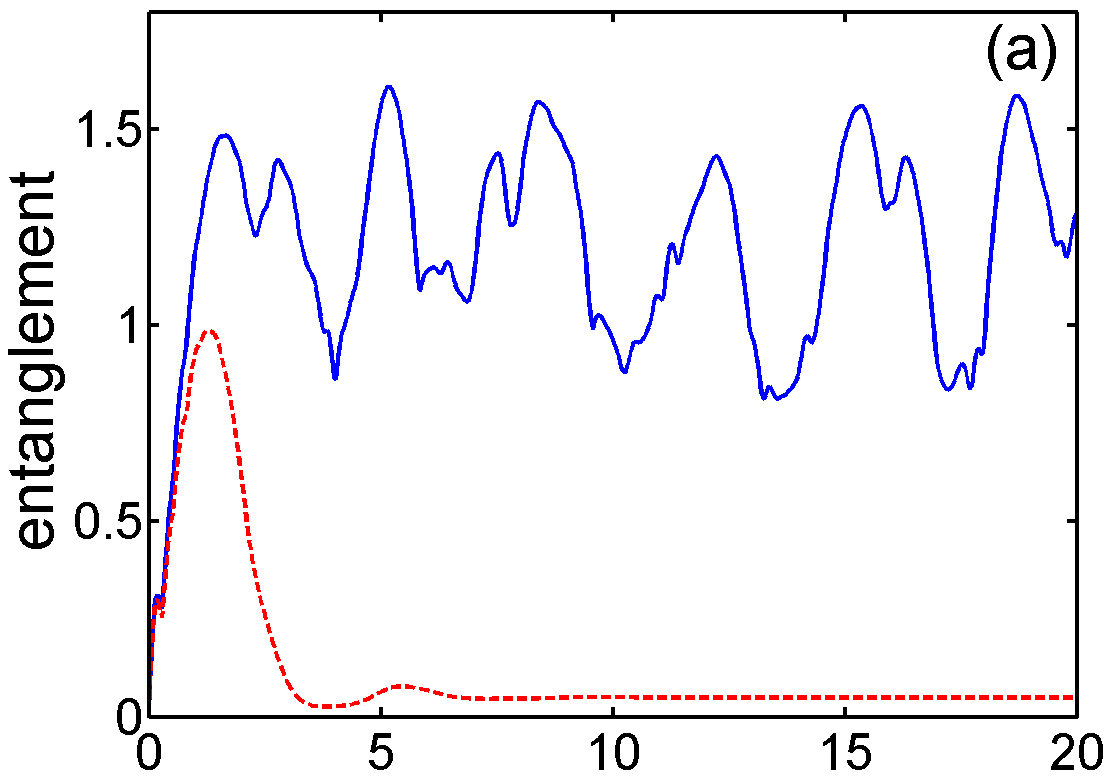}

\hspace*{4pt}\includegraphics[width=6cm]{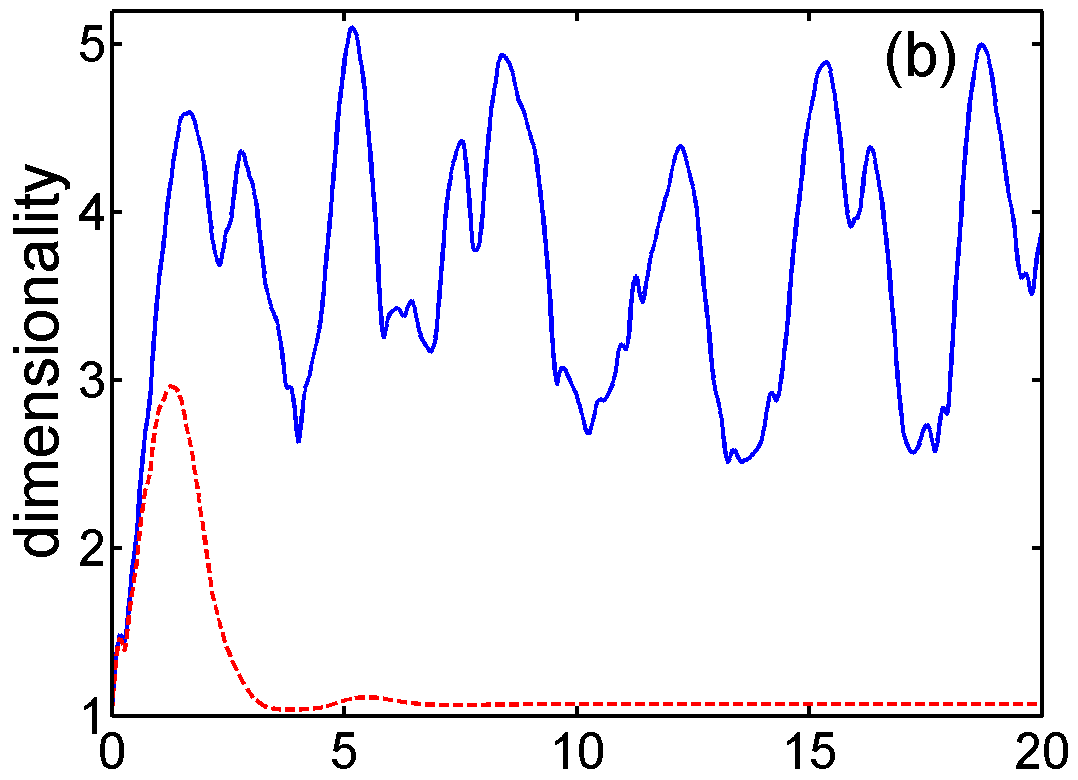}

\includegraphics[width=6.2cm]{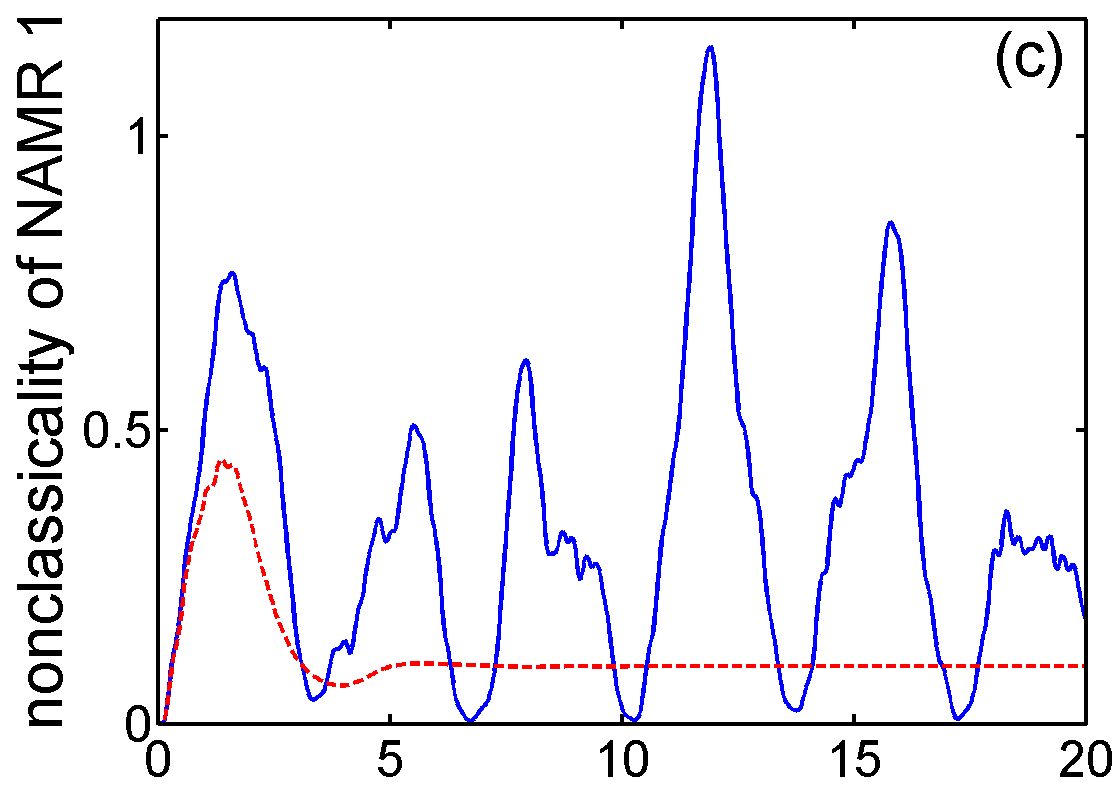}

\hspace*{4pt}\includegraphics[width=6cm]{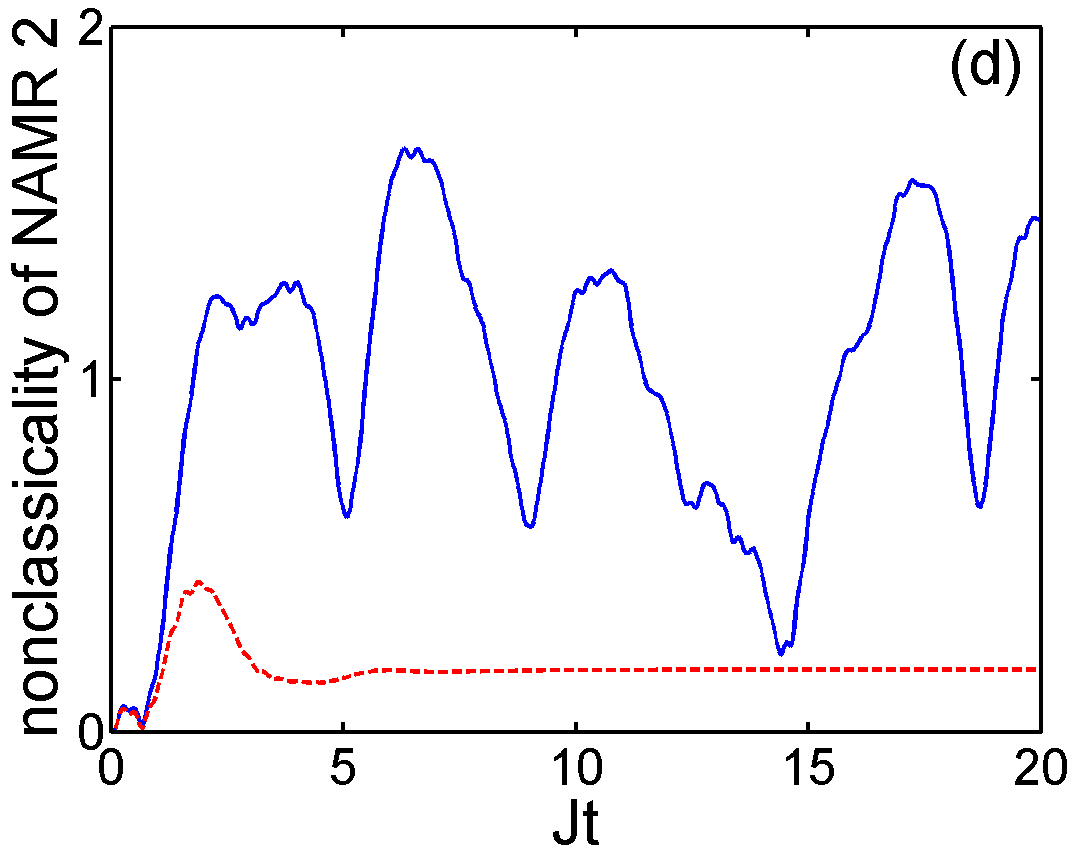}

\caption{(Color online) Non-stationary solutions for model~2: Same
as in Fig.~\ref{fig13} but for the effective Hamiltonian $H''_{\rm
eff}$. Additionally, panel (d) shows the nonclassicality, measured
by the entanglement potential $\EP(\rho^{(2)})$, of the second
NAMR. } \label{fig14}
\end{figure}

The evolutions of the nonclassicality of single NAMRs are plotted
in Fig.~\ref{fig13}(c) for model~1 and Figs.~\ref{fig14}(c)
and~\ref{fig14}(d)  for model~2. We find the following nonzero
values of $\EP(\rho^{(n)}_{\rm ss})$ in the corresponding steady
states: $\EP(\rho^{(1)}_{\rm ss})=\EP(\rho^{(2)}_{\rm ss})=0.1126$
for model~1, while $\EP(\rho^{(1)}_{\rm ss})=0.1354$ for the first
NAMR and $\EP(\rho^{(1)}_{\rm ss})=0.1770$ for the second NAMR in
model~2.

Here, for brevity, we studied only the evolution of one
nonclassicality measure. In future work, it might be physically
interesting to compare it with the evolution of other
measures~\cite{Miran15} and witnesses~\cite{Miran10} of
nonclassicality.


\section{Discussion and conclusions}

We studied tunable phonon blockade, which can be intuitively
understood as follows: Any number of phonons can be generated in a
harmonic resonator. However, this is not possible in an anharmonic
resonator, which is characterized by nonlinear (nonequidistant)
energy levels. So, if the driving field is in resonance with the
transition between the two lowest levels (say $\ket{0}$ and
$\ket{1}$), then it is not in resonance with the transitions
between the other levels. Thus, single-phonon blockade can be
observed. We showed in detail that a higher-order $n$-phonon
blockade can also be observed in a dissipative nonlinear system if
the driving field is resonant with the transition between other
levels $\ket{0}$ and $\ket{n}$. By applying coupled nonlinear
systems, instead of a single system, one can more easily tune
various types of multi-phonon two-mode blockades, as studied in
detail in this paper.

It is important to clarify the main differences between our model
of coupled oscillators and that studied in, e.g.,
Refs.~\cite{Liew10,Bamba11,Didier11}: (1) Here we assumed that the
oscillators are nonlinearly coupled as described by the
Hamiltonian $H_{\rm int}\sim(a_1+a^\dagger_1)(a_2+a^\dagger_2)$ in
contrast to the linear coupling given by $H_{\rm
int}\sim(a^\dagger_1 a_2+a_1 a^\dagger_2)$, which was applied in
Refs.~\cite{Liew10,Bamba11,Didier11}. Note that this linear
coupler (which is formally equivalent to a beam splitter or a
frequency converter), does not change the nonclassicality of a
total system~\cite{Miran15}. In contrast to this, the nonlinear
coupler can increase the nonclassicality of a total system, as
measured by, e.g., the entanglement potential. It is also worth
noting that, by applying our precise numerical calculations, we
found the steady-states of the NAMRs to be only partially coherent
(partially mixed), while the steady states calculated in, e.g.,
Ref.~\cite{Bamba11} were assumed to be completely coherent
(perfectly pure). (2) We have derived an effective Kerr-type
Hamiltonian from a microscopic one, given in Eq.~(\ref{N0}), while
an analogous Kerr-type Hamiltonian in
Refs.~\cite{Liew10,Bamba11,Didier11} was assumed without
derivation. Moreover, we studied here the Kerr interaction under
two different resonance conditions, as described by
Eqs.~(\ref{Hkerr1}) and~(\ref{Hkerr2}).
References~\cite{Liew10,Bamba11,Didier11} discussed only the Kerr
interaction given by Eq.~(\ref{Hkerr1}). (3) We analyzed the
blockade of mechanical phonons, contrary to photon blockade
studied in Refs.~\cite{Liew10,Bamba11}. Moreover, the interplay
between single-phonon blockade in one oscillator and single-photon
blockade in another oscillator was studied in
Ref.~\cite{Didier11}. Here we predicted tunable $k$-phonon
blockades (with $k=1,2,3$) in each oscillator, where the
$k$-phonon Fock state impedes the excitation of more phonons. To
our knowledge, multiphonon blockade has not been studied before.

In conclusion, we showed here a rich tapestry of phonon blockade
effects in two coupled nonlinear nanomechanical resonators.
Different types of phonon blockade could be ``picked out'' of this
tapestry by controlling the nonlinearity via ancilla TLS, and by
changing the driving frequency of the resonators themselves.
Within these different types of phonon blockade, the coupled NAMRs
can be made to behave like two coupled qubits, a qutrit coupled to
a quartit, or even two coupled qudits. We verified this picture by
looking at the nonclassical properties of these states including
their single-NAMR nonclassicality, and two-NAMR entanglement and
entanglement dimensionality. The nonclassical properties of these
states were also analyzed in phase space by applying the
$s$-parametrized Cahill-Glauber quasiprobability distributions
and, in particular, the Wigner function. We expect that, if
realized in experiment, the ability to operate in these different
regimes will have a range of applications in quantum information
and quantum technologies.

\section*{Acknowledgement}

A.M. was supported by the Polish National Science Centre under the
grants No. DEC-2011/02/A/ST2/00305 and No.
DEC-2011/03/B/ST2/01903. A.M. acknowledges a long-term fellowship
from the Japan Society for the Promotion of Science (JSPS). Y.X.L.
was supported by the National Natural Science Foundation of China
under Grant No. 61328502, the National Basic Research Program of
China (973 Program) under Grant No. 2014CB921401, the Tsinghua
University Initiative Scientific Research Program, and the
Tsinghua National Laboratory for Information Science and
Technology (TNList) Cross-discipline Foundation. J.B. was
supported by the Palack\'y University under the Project No.
IGA-P\v{r}F-2015-002. F.N. was partially supported by the RIKEN
iTHES Project, the MURI Center for Dynamic Magneto-Optics via the
AFOSR award number FA9550-14-1-0040, the IMPACT program of JST,
and a Grant-in-Aid for Scientific Research (A).

\appendix

\section{Derivation of the effective Hamiltonian}

Here we show how to derive the effective Hamiltonian $H_{\rm
eff}$, given by Eq.~(\ref{Heff}), from the Hamiltonian $H'''$,
given in Eq.~(\ref{H3}). The latter can be divided into the
following parts:
\begin{equation}
  H''' = H'''_{\rm sys} + H'''_{\rm drv} + H'''_{\rm int},
 \label{N29}
\end{equation}
where
\begin{eqnarray}
  H'''_{\rm sys} &=&H'''_0+H'''_{\rm JC}
  = \sum_n \Delta_n a_n^\dagger a_n +H'''_{\rm JC},
  \label{N27}\\
  H'''_{\rm drv}&=&\sum_n f_n(a+a^\dagger).
\label{N25}
\end{eqnarray}
Here $H'''_{\rm JC}$ is given by Eq.~(\ref{H_JC3}) and $H'''_{\rm
int}=H'_{\rm int}$ is given by Eq.~(\ref{Hint1}).

Our derivation is based on the method described in
Ref.~\cite{BGB09} for the exact diagonalization of the
Jaynes-Cummings model with the following unitary transformation
\begin{equation}
 U^{}_n= \exp[-\Lambda(\lambda_n)(a^\dagger_n \rho^-_n-a_n \rho^+_n)],
 \label{N22}
\end{equation}
where the operator $\Lambda(\lambda_n)=-\arctan(2\lambda_n
\sqrt{N_n})/(2\sqrt{N_n})$ is given in terms of the total number
of excitations $N_n=a^\dagger_n a_n+\ket{E_n}\bra{E_n}$ in the
$n$th NAMR and TLS. Thus, the expansions of the annihilation
operators, $\bar{a}_n=U_n^\dagger a U_n$ and
$\bar\rho_n^{-}=U_n^\dagger \rho_n^{-} U_n$ of Ref.~\cite{BGB09}
can be rewritten for our dressed qubit states as follows:
\begin{eqnarray}
  \bar a_n = a_n (1+\tfrac12 \lambda_n^2 \rho_n^z)+\lambda_n h_n^{(3)}
  \rho_n^{-}+\lambda_n^3 a_n^2 \rho_n^{+} + {\cal O}(\lambda^4_n),\;\;
 \label{N44} \\
 \bar\rho_n^{-} = h_n^{(1)} \rho_n^{-}+ \lambda_n a_n \rho_n^{z}
 - \lambda_n^2 a_n^2 \rho_n^{+} + {\cal O}(\lambda^3_n),\qquad
  \label{N45}
\end{eqnarray}
where $h_n^{(k)}=1-k \lambda_n^2 (a_n^\dagger a_n+1/2)$ and ${\cal
O}(\lambda^k_n)$ denotes the omitted terms of order
$\sim\lambda^k_n$ and higher. Now one can easily transform the
Hamiltonian $H'''$ into
\begin{eqnarray}
 \bar H &=& U_1^\dagger U_2^\dagger H''' U_2 U_1.
\label{N24}
\end{eqnarray}
In particular, by applying Eq.~(\ref{N44}), $H'''_{\rm drv}$ transforms into
\begin{equation}
  \bar H_{\rm drv}= \sum_n f_n[\bar a_n+\bar a^\dagger].
 \label{N26}
\end{equation}
If the qubits remain in the excited
dressed-qubit states $\ket{E_n}$, given in Eq.~(\ref{EG}), then
\begin{equation}
  \bra{E_1E_2}\bar H_{\rm drv}\ket{E_1E_2}
  = \sum_n f_n(a_n+a_n^\dagger)
  (1+\lambda_n^2/2)+ {\cal O}(\lambda^4_n).
 \label{N37}
\end{equation}
The assumption of a ``frozen''  state of  both qubits is
physically justified for the large detuning $\Delta_{\rm
rq}^{(n)}\gg g_n$, as specified in Eq.~(\ref{detuning1}).
Moreover, $H'''_{\rm int}$ transforms into
\begin{equation}
\bar H_{\rm int}= J_{12}(\bar a_{1} +\bar a_{1}^{\dagger})(\bar
a_{2}+\bar a_{2}^{\dagger}),
 \label{N39a}
\end{equation}
and, thus,
\begin{equation}
  \bra{E_1E_2}\bar H_{\rm int}\ket{E_1E_2}
  = J(a_{1}  +a_{1}^{\dagger})(a_{2}+a_{2}^{\dagger} )+ {\cal O}(\lambda^4_n),
 \label{N32}
\end{equation}
where $J=J_{12}(1+\lambda_1^2/2)(1+\lambda_2^2/2)$. Analogously,
by generalizing the results of Ref.~\cite{BGB09} for the
dressed-qubit operators $\rho_n^z$, one can find that
\begin{eqnarray}
  \bar H_{\rm sys} &=& U^{\dagger }H'''_{\rm sys}U
  -i U^\dagger \frac{\partial}{\partial t} U \label{N33a}\hspace{3cm} \\
&=& H'''_{0}
  - \tfrac12 \sum_n \delta_n \big(1-\sqrt{1+4N_{n}\lambda^2_n}\big)\rho_n^z
\hspace{1.7cm}\nonumber\\
  &\approx& H'''_{0} + \sum_n 2K_n a_n^\dagger a_n +
   h_n\rho_n^z +K_n (a_n^\dagger )^{2}a_n^{2}\rho_n^z,
\nonumber
\end{eqnarray}
where $H'''_0$ is defined in Eq.~(\ref{N27}),
$h_n=\chi_n(a_n^\dagger a_n +\tfrac12)$, with
$\chi_{n}=g'_n\lambda_n(1-\lambda_n^2)$, and the effective Kerr
nonlinearity reads $K_n=-g'_n\lambda^3_n =-(g'_n)^4/\delta_n^3$.
Moreover, $U=U_1U_2$ and $ \frac{\partial}{\partial t} U=0$. Thus,
one can write\begin{eqnarray}
  \bra{E_1E_2}\bar H_{\rm sys}\ket{E_1E_2}
 &=& \sum_n C_n + D_n a_n^\dagger a_n \nonumber \\ &&+ K_{n} (a_n^\dagger)^2
 a^2_n+ {\cal O}(\lambda^4_n),\quad
 \label{N33}
\end{eqnarray}
where $C_n=\chi_n/2$ (which can be neglected as a constant term)
and $D_n\equiv {\cal E}_n -\omega_{\rm
drv}^{(n)}=\Delta_n+2K_n+\chi_n$. Finally, the effective
Hamiltonian
\begin{equation}
  H_{\rm eff} \equiv \bra{E_1E_2} (\bar H_{\rm sys} + \bar H_{\rm drv} + \bar H_{\rm int})\ket{E_1E_2},
 \label{N41}
\end{equation}
is given explicitly by Eq.~(\ref{Heff}), where the terms $\sim
{\cal O}(\lambda^4_n)$ are omitted.

\section{Probability amplitudes in Eqs.~(\ref{sol1}) and~(\ref{sol2})}

The probability amplitudes $c_{xy}(t)=\langle xy|\psi(t)\rangle$
(for $x,y=0,...,3$), given in Eqs.~(\ref{sol1}) and~(\ref{sol2}),
can be obtained using the eigenvalue decompositions  $H'_{\rm
eff}\ket{E'_n}=E'_n\ket{E'_n}$, as
\begin{equation}
  c_{xy}(t)= \sum_n \exp(-iE'_nt) \langle E'_n|00\rangle \langle
  xy|E'_n\rangle,
 \label{cxy}
\end{equation}
and analogously for $H''_{\rm eff}\ket{E''_n}=E''_n\ket{E''_n}$.
To simplify these problems, let us limit the dimension of the
Hilbert space to that of two qutrits and assume $K_1=K_2=10J$ and
$J=F_1=F_2$ (as in Figs.~\ref{fig07} and~\ref{fig08}). Then the
Hamiltonians, given in Eqs.~(\ref{Heff1}) and~(\ref{Heff2}),
reduce to
\begin{equation}
\frac{H'_{\rm
eff}}{J}= \left(
\begin{array}{ccccccccc}
 0 & 1 & 0 & 1 & 1 & 0 & 0 & 0 & 0 \\
 1 & 0 & \sqrt{2} & 1 & 1 & \sqrt{2} & 0 & 0 & 0 \\
 0 & \sqrt{2} & 20 & 0 & \sqrt{2} & 1 & 0 & 0 & 0 \\
 1 & 1 & 0 & 0 & 1 & 0 & \sqrt{2} & \sqrt{2} & 0 \\
 1 & 1 & \sqrt{2} & 1 & 0 & \sqrt{2} & \sqrt{2} & \sqrt{2} & 2 \\
 0 & \sqrt{2} & 1 & 0 & \sqrt{2} & 20 & 0 & 2 & \sqrt{2} \\
 0 & 0 & 0 & \sqrt{2} & \sqrt{2} & 0 & 20 & 1 & 0 \\
 0 & 0 & 0 & \sqrt{2} & \sqrt{2} & 2 & 1 & 20 & \sqrt{2} \\
 0 & 0 & 0 & 0 & 2 & \sqrt{2} & 0 & \sqrt{2} & 40 \\
\end{array}
\right) ,
 \label{Hspecial1}
\end{equation}
\begin{equation}
\frac{H''_{\rm eff}}{J}= \left(
\begin{array}{ccccccccc}
 20 & 1 & 0 & 1 & 1 & 0 & 0 & 0 & 0 \\
 1 & 0 & \sqrt{2} & 1 & 1 & \sqrt{2} & 0 & 0 & 0 \\
 0 & \sqrt{2} & 0 & 0 & \sqrt{2} & 1 & 0 & 0 & 0 \\
 1 & 1 & 0 & 20 & 1 & 0 & \sqrt{2} & \sqrt{2} & 0 \\
 1 & 1 & \sqrt{2} & 1 & 0 & \sqrt{2} & \sqrt{2} & \sqrt{2} & 2 \\
 0 & \sqrt{2} & 1 & 0 & \sqrt{2} & 0 & 0 & 2 & \sqrt{2} \\
 0 & 0 & 0 & \sqrt{2} & \sqrt{2} & 0 & 40 & 1 & 0 \\
 0 & 0 & 0 & \sqrt{2} & \sqrt{2} & 2 & 1 & 20 & \sqrt{2} \\
 0 & 0 & 0 & 0 & 2 & \sqrt{2} & 0 & \sqrt{2} & 20 \\
\end{array}
\right),
 \label{Hspecial2}
\end{equation}
respectively. It is seen that the matrices, given by
Eqs.~(\ref{Hspecial1}) and~(\ref{Hspecial2}), differ only in their
diagonal terms. Unfortunately, even in these special cases, it is
very unlikely that exact analytical compact-form solutions of
these eigenvalue problems can be found, as they require finding
the roots of sixth and ninth order equations, respectively.
Especially, highly-irregular oscillations of $c_{xy}(t)$ for
model~2, as shown in Fig.~\ref{fig08}, confirm this conclusion.



\begin{thebibliography}{23}
\bibitem{Huang03}
X.~M.~H.~Huang, C.~A.~Zorman,  M.~Mehregany, and M.~L.~Roukes,
\extra{Nanoelectromechanical systems: {N}anodevice motion at
microwave frequencies,} Nature (London) \textbf{421}, 496 (2003).


\bibitem{Knobel03}
R.~G.~Knobel and A.~N.~Cleland, \extra{Nanometre-scale
displacement sensing using a single electron transistor,} Nature
(London) \textbf{424}, 291 (2003).


\bibitem{Blencowe04} M.~P.~Blencowe, \extra{Nanomechanical Quantum
Limits,} Science \textbf{304}, 56 (2004).


\bibitem{LaHaye04}
M.~D.~LaHaye, O.~Buu, B.~Camarota, and K.~C.~Schwab,
\extra{Approaching the quantum limit of a nanomechanical
resonator,} Science \textbf{304}, 74 (2004).


\bibitem{Blencowe04review}
M.~P.~Blencowe, \extra{Quantum electromechanical systems,} Phys.
Rep. \textbf{395}, 159 (2004).


\bibitem{Schwab05review}
K. C. Schwab and M. L. Roukes, \extra{Putting Mechanics into
Quantum Mechanics,} Phys. Today {\bf 58} (7), 36, (2005).


\bibitem{Ekinci05review}
K. L. Ekinci and M. L. Roukes, \extra{Nanoelectromechanical
systems,} Rev. Sc. Inst.~\textbf{76}, 061101 (2005).


\bibitem{Aspelmeyer14review}
M. Aspelmeyer, T. J. Kippenberg, and F. Marquardt, \extra{Cavity
Optomechanics,} Rev. Mod. Phys. \textbf{86}, 1391 (2014).


\bibitem{Caves80}
C. Caves, K. Thorne, R. Drever, V. D. Sandberg, and M. Zimmermann,
\extra{On the measurement of a weak classical force coupled to a
quantum-mechanical oscillator,} Rev. Mod. Phys. {\bf 52}, 341
(1980).


\bibitem{Bocko96}
M.~F.~Bocko and R.~Onofrio, \extra{On the measurement of a weak
classical force coupled to a harmonic oscillator: {E}xperimental
progress,} Rev. Mod. Phys.~\textbf{68}, 755 (1996).


\bibitem{Buks06}
E.~Buks and B.~Yurke, \extra{Mass detection with a nonlinear
nanomechanical resonator,} Phys. Rev. E~\textbf{74}, 046619
(2006).


\bibitem{Braginsky92}
V.~B.~Braginsky and F.~Ya.~Khalili, \textit{Quantum Measurements}
(Cambridge University Press, Cambridge, England, 1992).


\bibitem{OConnell10}
A. D. O'Connell \emph{et al.}, \extra{Quantum ground state and
single-phonon control of a mechanical resonator,} Nature (London)
\textbf{464}, 697 (2010).


\bibitem{ClelandBook}
A.~N.~Cleland, \textit{Foundations of Nanomechanics} (Springer,
Berlin, 2003).


\bibitem{Teufel11}
J. D. Teufel, D. Li, M. S. Allman, K. Cicak, A. J. Sirois, J. D.
Whittaker, and R. W. Simmonds, \extra{Circuit cavity
electromechanics in the strong-coupling regime,} Nature (London)
\textbf{471}, 204 (2011).


\bibitem{Chan11}
J. Chan, T.P. Mayer-Alegre, A. H. Safavi-Naeini, J. T. Hill, A.
Krause, S. Gr\"oblacher, M. Aspelmeyer, and O. Painter,
\extra{Laser cooling of a nanomechanical oscillator into its
quantum ground state,} Nature (London) \textbf{478}, 89 (2011).


\bibitem{Safavi12}
A. H. Safavi Naeini, J. Chan, J. T. Hill, T. P. Mayer Alegre, A.
Krause, and O. Painter, \extra{Observation of Quantum Motion of a
Nanomechanical Resonator,} Phys. Rev. Lett. \textbf{108}, 033602
(2012).


\bibitem{Verhagen12}
E. Verhagen, S. Deleglise, S. Weis, A. Schliesser, and T. J.
Kippenberg, \extra{Quantum-coherent coupling of a mechanical
oscillator to an optical cavity mode,} Nature (London)
\textbf{482}, 63 (2012).


\bibitem{Stannigel10}
K. Stannigel, P. Rabl, A.S. S\o{}rensen, P. Zoller, and M.D.
Lukin, \extra{Optomechanical transducers for long-distance quantum
communication,} Phys. Rev. Lett. {\bf 105}, 220501 (2010).


\bibitem{Palomaki13}
T. A. Palomaki, J. W. Harlow, J. D. Teufel, R. W. Simmonds, and K.
W. Lehnert, \extra{Coherent state transfer between itinerant
microwave fields and a mechanical oscillator,} Nature (London)
\textbf{495}, 210 (2013).


\bibitem{Huang13}
P. Huang, P. Wang, J. Zhou, Z. Wang, C. Ju, Z. Wang, Y. Shen, C.
Duan, and J. Du, \extra{Demonstration of Motion Transduction Based
on Parametrically Coupled Mechanical Resonators,} \prl {\bf 110},
227202 (2013).


\bibitem{Fu14}
H. Fu, T. H. Mao, Y. Li, J. F. Ding, J. D. Li, and G. Cao,
\extra{Optically mediated spatial localization of collective modes
of two coupled cantilevers for high sensitivity optomechanical
transducer,} Appl. Phys. Lett. {\bf 105}, 014108 (2014).


\bibitem{Wallquist09}
M. Wallquist, K. Hammerer, P. Rabl, M. Lukin, and P. Zoller,
\extra{Hybrid quantum devices and quantum engineering,} Phys. Scr.
{\bf T 137}, 014001 (2009).


\bibitem{Safavi11}
A. H. Safavi-Naeini and O. Painter, \extra{Proposal for an
optomechanical traveling wave phonon-photon translator,} New J.
Phys. {\bf 13}, 013017 (2011).


\bibitem{Regal08}
C. A. Regal, J. D. Teufel, and  K. W. Lehnert, \extra{Measuring
nanomechanical motion with a microwave cavity interferometer,}
Nat. Phys. {\bf 4}, 555 (2008).


\bibitem{Woolley08}
M.~J. Woolley, G.~J. Milburn, and C.~M. Caves, \extra{Nonlinear
quantum metrology using coupled nanomechanical resonators,} New J.
Phys. {\bf 10}, 125018 (2008).


\bibitem{Tian92}
L. Tian and H. J. Carmichael, \extra{Quantum trajectory
simulations of two-state behavior in an optical cavity containing
one atom,} \pra \textbf{46}, R6801 (1992).


\bibitem{Leonski94}
W. Leo\'nski and R. Tana\'s, \extra{Possibility of producing the
one-photon state in a kicked cavity with a nonlinear Kerr medium,}
\pra \textbf{49}, R20 (1994).


\bibitem{Miran96}
A. Miranowicz, W. Leo\'nski, S. Dyrting, and R. Tana\'s,
\Title{Quantum state engineering in finite-dimensional Hilbert
space,} Acta Phys. Slovaca  \textbf{46}, 451 (1996).


\bibitem{Imamoglu97}
A.~Imamo\={g}lu, H.~Schmidt, G. Woods, and M. Deutsch,
\extra{Strongly Interacting Photons in a Nonlinear Cavity,} \prl
\textbf{79}, 1467 (1997).


\bibitem{Werner99} M. J. Werner and A. Imamo\={g}lu,
\extra{Photon-photon interactions in cavity electromagnetically
induced transparency,} \pra \textbf{61}, 011801 (1999).


\bibitem{Brecha99}
R. J. Brecha, P. R. Rice, and M. Xiao, \extra{N two-level atoms in
a driven optical cavity: {Q}uantum dynamics of forward photon
scattering for weak incident fields,} \pra \textbf{59}, 2392
(1999).


\bibitem{Rebic99} S. Rebi\'{c}, S. M. Tan, A. S. Parkins, and D. F.
Walls, \extra{Large Kerr nonlinearity with a single atom,} J. Opt.
B \textbf{1}, 490 (1999).


\bibitem{Kim99} J. Kim, O. Bensen, H. Kan, and Y. Yamamoto,
\extra{A single-photon turnstile device,} Nature (London)
\textbf{397}, 500 (1999).


\bibitem{Rebic02} S. Rebi\'c, A. S. Parkins, and S. M. Tan,
\extra{Photon statistics of a single-atom intracavity system
involving electromagnetically induced transparency,}  \pra
\textbf{65}, 063804 (2002).


\bibitem{Smolyaninov02} I. I. Smolyaninov, A. V. Zayats, A. Gungor, and C.
C. Davis, \extra{Single-Photon Tunneling via Localized Surface
Plasmons,} \prl \textbf{88}, 187402 (2002).


\bibitem{Hoffman11}
A. J. Hoffman, S. J. Srinivasan, S. Schmidt, L. Spietz, J.
Aumentado, H. E. Tureci, and A. A. Houck, \extra{Dispersive Photon
Blockade in a Superconducting Circuit,} \prl \textbf{107}, 053602
(2011).


\bibitem{Lang11}
C. Lang \etal, \extra{Observation of Resonant Photon Blockade at
Microwave Frequencies Using Correlation Function Measurements,}
\prl \textbf{106}, 243601 (2011).


\bibitem{Liu14}
Y.X. Liu, X.W. Xu, A. Miranowicz, and F. Nori, \extra{From
blockade to transparency: {C}ontrollable photon transmission
through a circuit QED system,} \pra \textbf{89}, 043818 (2014).


\bibitem{Miran01}
A. Miranowicz, W. Leo\'nski, and N. Imoto, \extra{Quantum-optical
states in finite-dimensional Hilbert space. I. General formalism,}
Adv. Chem. Phys. \textbf{119}, 155 (2001); W. Leo\'nski and A.
Miranowicz, \extra{Quantum-optical states in finite-dimensional
Hilbert space. II. State generation,} Adv. Chem. Phys.
\textbf{119}, 195 (2001).


\bibitem{Leonski11review}
W. Leo\'nski and A. Kowalewska-Kud{\l}aszyk, \extra{Quantum
scissors: {F}inite-dimensional states engineering,} in {\em
Progress in Optics,} edited by E. Wolf (Elsevier, Amsterdam,
2011), Vol. {\bf 56}, p. 131.


\bibitem{Birnbaum05}
K. M. Birnbaum, A. Boca, R. Miller, A. D. Boozer, T. E. Northup,
and H. J. Kimble, \extra{Photon blockade in an optical cavity with
one trapped atom,} Nature (London) \textbf{436}, 87 (2005).


\bibitem{Liu10}
Y. X. Liu, A. Miranowicz, Y. B. Gao, J. Bajer, C. P. Sun, and F.
Nori, \extra{Qubit-induced phonon blockade as a signature of
quantum behavior in nanomechanical resonators,} \pra \textbf{82},
032101 (2010).


\bibitem{Didier11}
N. Didier, S. Pugnetti, Y. M. Blanter, and R. Fazio,
\extra{Detecting phonon blockade with photons,} \prb \textbf{84},
054503 (2011).


\bibitem{Johansson14}
J.R. Johansson, N. Lambert, I. Mahboob, H. Yamaguchi, F. Nori,
\extra{Entangled-state generation and Bell inequality violations
in nanomechanical resonators,} Phys. Rev. B {\bf 90}, 174307
(2014).


\bibitem{Buluta11}
I. Buluta, S. Ashhab, and F. Nori, \extra{Natural and artificial
atoms for quantum computation,} Rep. Prog. Phys. \textbf{74},
104401 (2011).


\bibitem{Palomaki13b}
T. A. Palomaki, J. D. Teufel, R. W. Simmonds, and K. W. Lehnert,
\extra{Entangling Mechanical Motion with Microwave Fields},
Science {\bf 342}, 710 (2013).


\bibitem{Okamoto13}
H. Okamoto, A. Gourgout, C. Y. Chang, K. Onomitsu, I. Mahboob, E.
Y. Chang, and H. Yamaguchi, \extra{Coherent phonon manipulation in
coupled mechanical resonators,} Nat. Phys. {\bf 9}, 480 (2013).


\bibitem{Cleland04}
A.~N.~Cleland and M.~R.~Geller, \extra{Superconducting Qubit
Storage and Entanglement with Nanomechanical Resonators,}
\prl~{\bf 93}, 070501 (2004).


\bibitem{Armour02} A.~D.~Armour, M.~P.~Blencowe, and K.~C.~Schwab,
\extra{Entanglement and Decoherence of a Micromechanical Resonator
via Coupling to a Cooper-Pair Box,} \prl~\textbf{88}, 148301
(2002).


\bibitem{Tian05}
L.~Tian, \extra{Entanglement from a nanomechanical resonator
weakly coupled to a single Cooper-pair box,} \prb~\textbf{72},
195411 (2005).


\bibitem{Jacobs09}
K. Jacobs and A. J. Landahl, \extra{Engineering Giant
Nonlinearities in Quantum Nanosystems,} \prl \textbf{103}, 067201
(2009).


\bibitem{Hartman13}
S. Rips, and M. J. Hartmann, \extra{Quantum Information Processing
with Nanomechanical Qubits,} \prl  \textbf{110}, 120503, (2013).


\bibitem{Mahboob14}
I. Mahboob, H. Okamoto, K. Onomitsu, and H. Yamaguchi,
\extra{Two-Mode Thermal-Noise Squeezing in an Electromechanical
Resonator,} \prl {\bf 113}, 167203 (2014).


\bibitem{Miran13}
A. Miranowicz, M. Paprzycka, Y.X. Liu, J. Bajer, and  F. Nori,
\extra{Two-photon and three-photon blockades in driven nonlinear
systems,} \pra~\textbf{87}, 023809 (2013).


\bibitem{Hovsepyan14}
G.H. Hovsepyan, A.R. Shahinyan, and G.Y. Kryuchkyan,
\extra{Multiphoton blockades in pulsed regimes beyond stationary
limits,} \pra~\textbf{90}, 013839 (2014).


\bibitem{Miran14qre}
A. Miranowicz, J. Bajer, M. Paprzycka, Y. X. Liu, A. M. Zagoskin,
and F. Nori, \extra{State-dependent photon blockade via
quantum-reservoir engineering,} Phys. Rev. A \textbf{90}, 033831
(2014).

\bibitem{Wang15}
H. Wang, X. Gu, Y. X. Liu, A. Miranowicz, and F. Nori,
\extra{Tunable photon blockade in a hybrid system consisting of an
optomechanical device coupled to a two-level system,} \pra
\textbf{92}, 033806 (2015).

\bibitem{Faraon08}
A. Faraon, I. Fushman, D. Englund, N. Stoltz, P. Petroff, and J.
Vu\v{c}kovi\'c, \extra{Coherent generation of non-classical light
on a chip via photon-induced tunnelling and blockade,} Nat. Phys.
{\bf 4}, 859 (2008).


\bibitem{Majumdar12}
A. Majumdar, M. Bajcsy, and J. Vu\v{c}kovi\'c, \extra{Probing the
ladder of dressed states and nonclassical light generation in
quantum-dot-cavity QED,} \pra \textbf{85}, 041801(R) (2012).

\bibitem{CohenBook}
C. Cohen-Tannoudji, J. Dupont-Roc, and G. Grynberg, {\em
Atom-Photon Interactions} (Wiley, New York, 1992).


\bibitem{Liu06}
Y. X. Liu, C. P. Sun, and F. Nori, \extra{Scalable superconducting
qubit circuits using dressed states,} \pra \textbf{74}, 052321
(2006).


\bibitem{Leonski04coupler}
W. Leo\'nski and A. Miranowicz, \extra{Kerr nonlinear coupler and
entanglement,} \job~\textbf{6}, S37 (2004).


\bibitem{Liew10}
T. C. H. Liew and V. Savona, \extra{Single Photons from Coupled
Quantum Modes,} \prl \textbf{104}, 183601 (2010).


\bibitem{Bamba11}
M. Bamba, A. Imamoglu, I. Carusotto, and C. Ciuti, \extra{Origin
of strong photon antibunching in weakly nonlinear photonic
molecules,} \pra \textbf{83}, 021802(R) (2011).


\bibitem{Miran06coupler}
A. Miranowicz and W. Leo\'nski, \extra{Two-mode optical state
truncation and generation of maximally entangled states in pumped
nonlinear couplers,} J. Phys. B~\textbf{39}, 1683 (2006);


\bibitem{Xu14coupler}
X. W. Xu and Y. Li, \extra{Tunable photon statistics in weakly
nonlinear photonic molecules,} \pra \textbf{90}, 043822 (2014).


\bibitem{SzeTan99}
S. M. Tan, \extra{A computational toolbox for quantum and atomic
optics}, J. Opt. B: Quantum Semiclass. Opt. {\bf 1}, 424 (1999).

\bibitem{BreuerBook}
H. P. Breuer and F. Petruccione, \emph{The Theory of Open Quantum
Systems} (Oxford University Press, Oxford, 2003).

\bibitem{Miran16}
A. Miranowicz, N. Lambert, F. Nori \etal, in preparation.

\bibitem{Beaudoin11}
F. Beaudoin, J. M. Gambetta, and A. Blais, \extra{Dissipation and
ultrastrong coupling in circuit QED}, \pra 84, 043832 (2011).

\bibitem{Cahill69}
K. E. Cahill and R. J. Glauber, \extra{Ordered expansions in boson
amplitude operators,} Phys. Rev. \textbf{177}, 1857 (1969).


\bibitem{Miran15}
A. Miranowicz, K. Bartkiewicz, A. Pathak, J. Perina Jr., Y. N.
Chen, and F. Nori, \extra{Statistical mixtures of states can be
more quantum than their superpositions: {C}omparison of
nonclassicality measures for single-qubit states,} Phys. Rev. A
\textbf{91}, 042309 (2015); A. Miranowicz, K. Bartkiewicz, N.
Lambert, Y. N. Chen, and F. Nori, \extra{Increasing relative
nonclassicality quantified by standard entanglement potentials by
dissipation and unbalanced beam splitting,} Phys. Rev. A
\textbf{92}, 062314 (2015).


\bibitem{Horodecki09review}
R. Horodecki, P. Horodecki, M. Horodecki, and K. Horodecki,
\extra{Quantum entanglement,} Rev. Mod. Phys. {\bf 81}, 865
(2009).

\bibitem{Audenaert03}
K. Audenaert, M. B. Plenio, and J. Eisert, \extra{Entanglement
Cost under Positive-Partial-Transpose-Preserving Operations,} \prl
\textbf{90}, 027901 (2003).

\bibitem{Ishizaka04}
S. Ishizaka, \extra{Binegativity and geometry of entangled states
in two qubits,} \pra \textbf{69}, 020301(R) (2004).

\bibitem{Eltschka13}
C. Eltschka and J. Siewert, \extra{Negativity as an Estimator of
Entanglement Dimension,} \prl \textbf{111}, 100503 (2013).

\bibitem{Arkhipov15}
I. I. Arkhipov, J. Pe\v{r}ina Jr., J. Pe\v{r}ina, and A.
Miranowicz, \extra{Comparative study of nonclassicality,
entanglement, and dimensionality of multimode noisy twin beams,}
\pra \textbf{91}, 033837 (2015).

\bibitem{Asboth05}
J. K. Asboth, J. Calsamiglia, and H. Ritsch, \extra{Computable
Measure of Nonclassicality for Light,} \prl \textbf{94}, 173602
(2005).


\bibitem{Miran10}
A. Miranowicz, M. Bartkowiak, X. Wang, Y. X. Liu, and F. Nori,
\extra{Testing nonclassicality in multimode fields: {A} unified
derivation of classical inequalities,} \pra \textbf{82}, 013824
(2010); M. Bartkowiak, A. Miranowicz, X. Wang, Y.X. Liu, W.
Leo\'nski, and F. Nori, \extra{Sudden vanishing and reappearance
of nonclassical effects: {G}eneral occurrence of finite-time
decays and periodic vanishings of nonclassicality and entanglement
witnesses,} \pra \textbf{83}, 053814 (2011).


\bibitem{BGB09}
M. Boissonneault, J. M. Gambetta, and A. Blais, \extra{Dispersive
regime of circuit QED: {P}hoton-dependent qubit dephasing and
relaxation rates,} \pra \textbf{79}, 013819 (2009).

\end{thebibliography}
\end{document}